# Coexistence of Merons with Skyrmions in the Centrosymmetric van der Waals Ferromagnet Fe$_{5-x}$GeTe$_2$

*Brian W. Casas[†], Yue Li[†], Alex Moon, Yan Xin, Conor McKeever, Juan Macy, Amanda K. Petford-Long, Charudatta M. Phatak, Elton J. G. Santos, Eun Sang Choi, and Luis Balicas[*]*

([†]These authors contributed equally to this work.)

Fe$_{5-x}$GeTe$_2$ is a centrosymmetric, layered van der Waals (vdW) ferromagnet that displays Curie temperatures $T_c$ (270-330 K) that are within the useful range for spintronic applications. However, little is known about the interplay between its topological spin textures (e.g., merons, skyrmions) with technologically relevant transport properties such as the topological Hall effect (THE), or topological thermal transport. Here, we show via high-resolution Lorentz transmission electron microscopy that merons and anti-meron pairs coexist with Néel skyrmions in Fe$_{5-x}$GeTe$_2$ over a wide range of temperatures and probe their effects on thermal and electrical transport. We detect a THE, even at room $T$, that senses merons at higher $T$'s as well as their coexistence with skyrmions as $T$ is lowered indicating an on-demand thermally driven formation of either type of spin texture. Remarkably, we also observe an unconventional THE in absence of Lorentz force and attribute it to the interaction between charge carriers and magnetic field-induced chiral spin textures. Our results expose Fe$_{5-x}$GeTe$_2$ as a promising candidate for the development of applications in skyrmionics/meronics due to the interplay between distinct but coexisting topological magnetic textures and unconventional transport of charge/heat carriers.

Brian W. Casas, Alex Moon, Yan Xin, Juan Macy, Eun Sang Choi, and Luis Balicas[*]
National High Magnetic Field Laboratory, Tallahassee, 32310, Fl, USA.
E-mail: balicas@magnet.fsu.edu

Alex Moon, Juan Macy, Eun Sang Choi, and Luis Balicas[*]
Department of Physics, Florida State university, Tallahassee, 32310, Fl, USA.

Yue Li, Amanda K. Petford-Long, Charudatta M. Phatak
Materials Science Division, Argonne National Laboratory, Lemont, 60439, IL, USA.

Amanda K. Petford-Long, Charudatta M. Phatak

Department of Materials Science and Engineering, Northwestern University, Evanston, 60208, IL, USA.

Conor McKeever, Elton J. G. Santos

Institute for Condensed Matter and Complex Systems, School of Physics and Astronomy, The University of Edinburgh, EH9 3FD, UK.

Elton J. G. Santos

Higgs Centre for Theoretical Physics, The University of Edinburgh, EH9 3FD, UK.

## 1. Introduction

The emergence of layered two-dimensional (2D) magnets has opened a new domain of research focused not only on 2D magnetism[1] and complex topological spin textures[2], but also on moiré magnetism[3], novel heterostructure types[4], as well as spin- and valley-tronics[5]. Among vdW ferromagnets (FMs) $Fe_{5-x}GeTe_2$ has recently attracted substantial attention due to its relatively high Curie temperature[6] (~ 270 K and ~330 K[7]). $Fe_{5-x}GeTe_2$ crystallizes in a rhombohedral structure, with the space group *R-3m* and unit cell parameters *a* = 4.04(2) Å, and *c* = 29.19(3) Å[6]. The relatively large interlayer distance *c* results from three nonequivalent Fe sites that form four magnetic monolayers sandwiched among Te layers, including a honeycomb layer formed by two of the inequivalent Fe sites. Figure S1 in the Supporting Information shows transmission electron microscopy (TEM) images of a $Fe_{5-x}GeTe_2$ crystal exposing its structure. One of the inequivalent iron layers composed of the so-called Fe(1) site, located in the outermost $Fe_5Ge$ sublayer, is known to occupy one of two possible split-sites either above or below the Ge atom. A recent scanning tunneling microscopy study[8], shows evidence that the Fe(1) site orders in a $\sqrt{3} \times \sqrt{3}$ superstructure which seems to generate two coexisting phases with slightly different magnetic properties. This $\sqrt{3} \times \sqrt{3}$ ordering of the Fe(1)-Ge pair would break inversion symmetry and favor the antisymmetric exchange or Dzyaloshinskii-Moriya interaction (DMI). $Fe_{5-x}GeTe_2$ also undergoes a pronounced magneto-structural transition around $T_s$ ~ 115 K characterized by an abrupt reduction in the lattice constants *c*/*a* ratio and the emergence of new Bragg reflections[6].

Furthermore, the magneto-crystalline anisotropy in $Fe_{5-x}GeTe_2$ is weak and depends directly on the Fe deficiency $x$[9]. This low magnetic anisotropy associated with other magnetic interactions[10], including the presence of DMI, may induce the orientation of spins with either an easy-axis (perpendicular to the surface) or an easy-plane[6, 11]. Indeed, the coexistence, in the same crystal, of minor structural variations (Fe(1) arrangements) with different magnetic anisotropies is likely in $Fe_{5-x}GeTe_2$ due to their small difference in formation energies[17]. Such a complex structural arrangement liaised with competing exchange interactions may be behind the observation of different types of magnetic ground states and topological spin textures (e.g., skyrmions, merons)[2a, 12] in $Fe_{5-x}GeTe_2$. For example, it was proposed[8] that a peak in the magnetization seen just below $T_c$ at around 275 K, would correspond to the onset of helimagnetic order within the original ferromagnetic state. In fact, the competition between magnetic interactions is likely responsible for the observation[2a] of meron and anti-meron textures between magnetic domains in $Fe_{5-x}GeTe_2$ via Lorentz transmission electron microscopy (LTEM) under zero magnetic-field. A meron is a non-coplanar spin texture characterized by a quantum topological number $N = \pm1/2$. Merons are topologically equivalent to one-half of a skyrmion ($N = \pm1$) and in 2D ferromagnetic systems they exist only as pairs or within groups in 2D magnetic systems. However, a more recent LTEM study[13] on this compound reveals the existence of stripe-like, or labyrinthine Bloch domains, that would lead to the formation of Bloch like spin bubbles upon application of an external magnetic field. Bloch bubbles would be convertible into Néel skyrmions simply by decreasing the sample thickness[13]. Since skyrmions were observed in $Fe_{5-x}GeTe_2$ at lower temperatures and in exfoliated samples, it remains to be clarified whether merons and Néel skyrmions would coexist in this compound or if the proposed and observed magnetic phase-transitions would favor one type of spin texture in detriment of the other. In addition, if these spin textures have any effect on electrical and thermal transport properties for future real-device platforms is yet to be demonstrated.

Here, we report the coexistence of skyrmion and meron spin textures in $Fe_{5-x}GeTe_2$ ($x = \pm 0.4$) and their correlation with the thermal and topological Hall transport properties of the compound. We focus on $Fe_{5-x}GeTe_2$ because it differs in significant ways with respect to its more studied sister compound $Fe_{3-x}GeTe_2$: $R\text{-}3m$ structure with 3 inequivalent Fe sites[6] in contrast to $P6_3/mmc$ for the latter with 2 inequivalent sites, displays planar oriented moments in contrast to out-of-the plane ones, a significantly higher Curie temperature, i.e., up to $T_c \sim 330$ K[7] versus ~220 K, and a poorly understood magnetostructural transition at $T_s \sim 110$ K[6]. Here, we confirm the predominance of planar magnetic domains in the range of temperatures 170 K

≤ T ≤ 290 K but with magnetic vortices that can be identified, through micromagnetic simulations, as meron and anti-meron pairs. Between 100 K and 170 K, we observed the emergence of magnetic regions having a magneto-crystalline anisotropy oriented along the interlayer direction which host striped Néel-type magnetic domains. Additionally, Néel skyrmions are observed to nucleate under a field cooling process with the application of modest magnetic fields applied along the interlayer direction. The co-occurrence of both intralayer- and interlayer-oriented domains leads to the coexistence of merons and skyrmions for 100 K ≤ T ≤ 170 K. The presence and modulation of the spin textures are detected via a pronounced topological Hall-effect (THE) observed all the way up and beyond room temperature.

## 2. Results

All measurements displayed throughout this manuscript were collected in thermally cycled samples to suppress the structural metastability intrinsic to the first cool-down across the magnetostructural transition[6]. The conventional Hall-effect is measured by flowing an electrical current through a crystal (see Figure S1 for structural characterization of $Fe_{5-x}GeTe_2$ and Figure S2 for its magnetization as function of the temperature $T$) having a well-defined geometry and by placing leads at the edges of the sample to collect the Hall voltage induced by a magnetic field applied perpendicularly to the plane of the sample (see, schemes in **Figures 1a and 1b**). For a magnetic compound characterized by a pronounced spin orbit coupling, the Hall effect is observed to mimic the magnetization, displaying the anomalous Hall component: $\rho_{xy} = \rho_{xy}^N + S_H M \rho_{xx}^n$ where $\rho_{xy}^N$ is the conventional Hall response that depends on the density and mobility of charge carriers, $M$ is the magnetization of the sample, and $\rho_{xx}$ is its magnetoresistivity, with $n$ typically taking values close to $n \cong 2$. For fields applied along the interplanar direction and currents flowing within the conducting planes, $Fe_{5-x}GeTe_2$ displays an anomalous Hall response that roughly follows the magnetic field and the temperature dependence of the magnetization (Figures 1c, 1e). For fields beyond $\mu_0 H = 1$ T, $\rho_{xy}$ is observed to saturate at $T$-dependent values ranging between 3 and 6 μΩ cm. However, and as previously discussed in Ref.[2a], the Hall response in $Fe_{5-x}GeTe_2$ cannot be completely described in terms of the magnetization, magnetoresistivity, and the conventional Hall-effect.

As discussed through Figures S3 to S4, one can follow a careful procedure that considers the demagnetization factor of the sample, for this field orientation, to subtract the anomalous and conventional Hall components. The remanent, or the THE signal, yields a dip below $\mu_0 H \cong$ 0.5 T, whose magnitude ($\cong 1.2$ μΩ cm) is $T$-dependent. As we show below, this THE signal is observable over the entire temperature range including room temperature and above.

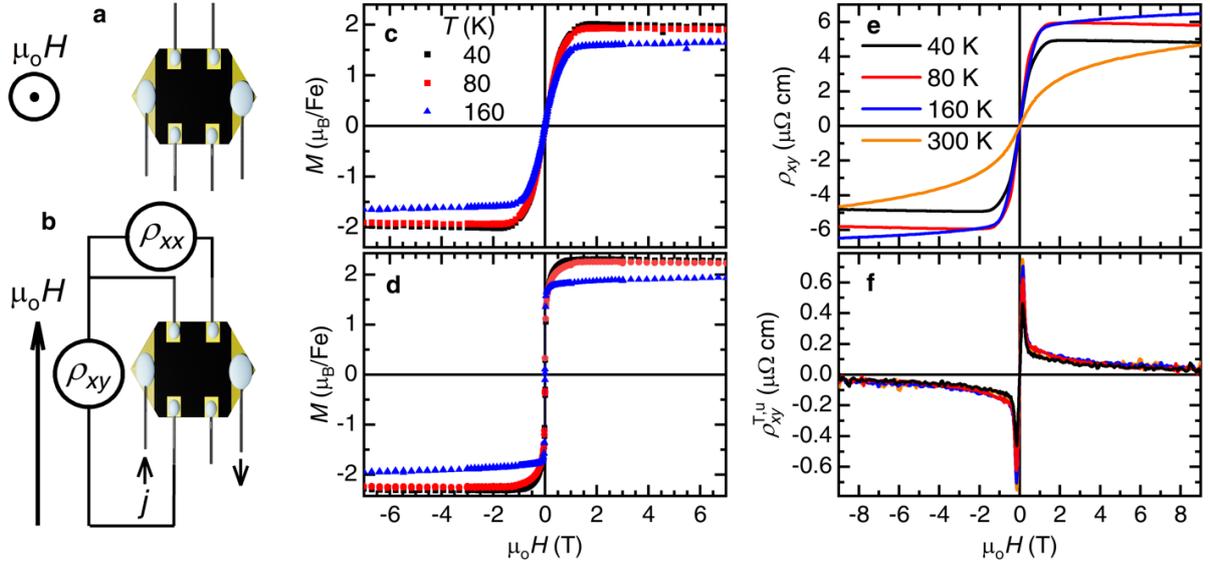

**Figure 1**. Anomalous and topological Hall effects in $Fe_{5-x}GeTe_2$. a-b) Configuration of measurements for collecting conventional and unconventional Hall responses, respectively. c-d) Magnetization $M$ as a function of magnetic field $\mu_0 H$ at different temperatures with the field oriented along the *c*-axis and the *ab*-plane, respectively. e, Raw conventional Hall response $\rho_{xy}$ for the same crystal, showing a clear anomalous Hall response that mimics the behavior of the magnetization as a function of both $\mu_0 H$ and $T$. f) Unconventional topological Hall-effect $\rho_{xy}^{T,u}$ (u-THE) extracted for electrical currents and $\mu_0 H$ aligned along the conducting planes. These traces were obtained after *anti-symmetrization* to subtract the superimposed magnetoresistivity. Notice the sharp peak observed at very low fields due to the u-THE. Here, the quoted magnetic field values for different sample orientations do not consider the contribution of the demagnetization factor intrinsic to the geometry of the sample.

Remarkably, when using a measurement scheme previously utilized to study the topological Hall effect in $Fe_{3-x}GeTe_2$[14] (Figure 1b), with the magnetic field oriented along a planar direction of the crystal, an antisymmetric or Hall-like response (Figure 1f) that does not reproduce the magnetization (Figure 1d) can be observed at all temperatures including those exceeding room *T*. This signal can also be observed when the electrical current flowing along the basal plane of the crystal is aligned along the external field, or in absence of Lorentz force. The important point is that we observe a true Hall like signal via an unconventional experimental configuration which a priori should yield none. An anomalous Hall response also in the absence of Lorentz force was reported for $ZrTe_5$[15]. Due to the unconventional nature of the experimental configuration used, and to distinguish it from the conventional topological

Hall-effect (in the presence of Lorentz force), we will denominate this signal as the unconventional topological signal (u-THE) or $\rho_{xy}^{T,u}$. Note the sharp asymmetric peak in $\rho_{xy}^{T,u}$ below $\mu_0 H = 1$ T (Figure 1f). The magnitude of which shows a significant temperature dependence, while the magnetic field at which the maximum response occurs appears insensitive to changes in temperature. This observation should not be confused with the so-called planar Hall-effect (PHE) discussed in the context of Weyl semi-metals[16] or anisotropic magnetic systems[17], which measures the anisotropy of the magnetoresistivity for fields rotating in the same plane of the electrical current. The PHE is an even magnetoresistivity signal obtained after averaging, or symmetrizing, negative and positive magnetic field sweeps. In contrast, $\rho_{xy}^{T,u}$ is the antisymmetric, or odd in magnetic field signal obtained after subtracting negative field sweep traces from positive ones, which behaves as a true Hall signal even in the absence of Lorentz force. This u-THE likely derives its existence from the deflection of charge carriers by the spin-chirality scalar field $S_{ijk} = \boldsymbol{S}_i \cdot (\boldsymbol{S}_j \times \boldsymbol{S}_k)$, intrinsic to topological spin textures such as skyrmions, merons, and possibly other non-coplanar spin textures either within the same labyrinthine FM domains or along their domain walls. It has been argued that the interaction between the itinerant carriers and topological spin textures is particularly strong in metallic systems where Hund's like coupling leads to the alignment of free carrier spins along the magnetic moments that participate in the topological textures[18]. Here, the important point is the observation of an unconventional topological Hall response above room temperature.

Figure S5 highlights the data from a 15 nm thick exfoliated crystal, encapsulated among $h$-BN layers. For this sample, we found it impossible to eliminate the high field $\rho_{xy}^{T,u}$ component, implying that it is intrinsic to the material and not an artifact from misalignment. In exfoliated samples the sharp peak evolves into a broad one emerging at and extending to much higher magnetic fields. This can only result from the evolution of the spin textures and domain structures upon exfoliation, akin to what was previously reported for Co doped Fe$_{5-x}$GeTe$_2$ samples[12b] that reveal skyrmions only within a precise range of sample thicknesses. Evidence for the role of exfoliation on spin textures and domain structure is provided by the observation of a large coercive magnetic field that is completely absent in bulk samples. In both exfoliated and bulk samples, the maxima observed in $\rho_{xy}^{T,u}$ display a clear temperature dependence with its maximum occurring at ~ 120 K, nearly coinciding with the reported value of $T_s$. A secondary maximum is observed around 240 K before the response begins to weaken due to the heightened effect of thermal fluctuations upon approaching $T_c$.

Given the layered and anisotropic nature of Fe$_{5-x}$GeTe$_2$, it is pertinent to ask if such unconventional THE response would replicate the conventional THE observed for fields aligned along the inter-planar direction[2a], given that one would naively expect the magnetic field to lead to distinct spin textures when applied along or perpendicularly to the conducting planes. The extraction of the c-THE requires the deconvolution of the normal Hall $\rho_{xy}^N$ and the anomalous Hall $\rho_{xy}^A$ responses from the measured Hall signal that was anti-symmetrized to remove the residual isothermal magnetoresistivity, $\rho_{xx}(\mu_0 H, T = \text{constant})$. The c-THE response[19], $\rho_{xy}^T(\mu_0 H_{\text{int}})$, which is observed for fields $\mu_0 H_{\text{int}} \lesssim 2$ T shows a peak whose amplitude is temperature dependent, reaching a maximum value in the vicinity of 160 K, (**Figure 2a**). The absolute maximum value of the c-THE response or $|\rho_{xy}^{T,\text{max}}|$ evolves nonmonotonically as a function of temperature (Figure 2b) and when associated to the required applied internal field for reaching its maximum, $\mu_0 H_{\text{int}}^{\text{max}}$, one can infer the existence of three distinct topological phases or regimes. Regime I is observed for $T \lesssim 80$ K, or below the magneto-structural transition at $T_s$[6, 20], where $\rho_{xy}^{T,\text{max}}$ shows a nearly monotonic increase upon warming towards 80 K (Figure 2b). The applied magnetic field $\mu_0 H_{\text{int}}^{\text{max}}$ of $\rho_{xy}^{T,\text{max}}$ increases slightly to a maximum value at 80 K, from a previously saturated value below 40 K (Figure 2b). This little to no invariance with respect to $T$ at the lowest temperatures, follows the behavior of the magnetization that decreases only slightly below 80 K. The anomalous Hall coefficient $S_H$ also remains nearly constant below 40 K (Figure 1c). Region II would correspond to temperatures 80 K ≤ $T$ ≤ 160 K, and region III to $T$ ≥ 160 K. Region II shows a maximum in $\rho_{xy}^T$ at 160 K with $\mu_0 H_{\text{int}}^{\text{max}}$ decreasing to less than half of its original value at 80 K. Finally, Region III shows a surprising decrease in the magnitude $\rho_{xy}^{T,\text{max}}$. As we will discuss below, based on our Lorentz-transmission microscopy study, regions II and III can be associated respectively, with the coexistence of Néel skyrmions and meron pairs due to rotation of the magneto-crystalline anisotropy upon approaching $T$s which stabilizes out-of-plane magnetic domains coexisting with in-plane ones, and the predominance of meron – anti-meron pairs at higher temperatures. Region I will remain for future LTEM studies, however a slightly weakened THE indicates the presence of chiral spin textures albeit affected by the magnetostructural transition. It is worth noting that the c-THE continues to have finite values at 300 K and is likely to do so beyond the Curie temperature at $T_c \simeq 310 - 330$ K depending on the precise Fe content.

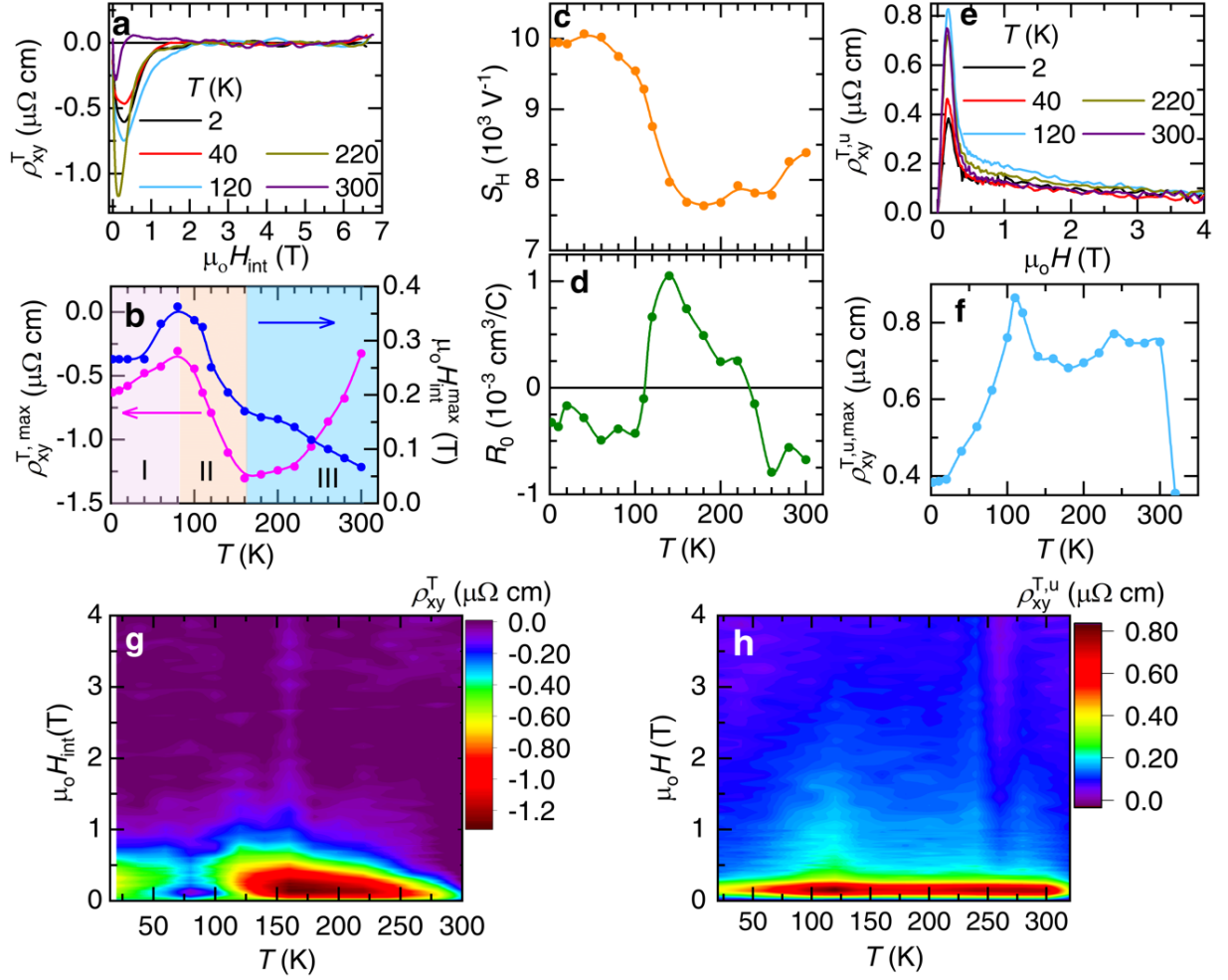

**Figure 2.** Conventional and unconventional topological Hall effects in $Fe_{5-x}GeTe_2$. a) Representative field dependence of the topological Hall effect component $\rho_{xy}^{T}$ for fields perpendicular to the basal plane at several temperatures. b) Amplitude of the maximum observed in the THE (magenta) $\rho_{xy}^{T,max}$, and the magnetic field value $\mu_0 H_{int}^{max}$ where the maximum occurs (blue) as a function of $T$. c-d) Anomalous Hall $S_H$ and conventional Hall $R_0$ coefficients as functions of $T$, respectively. $R_0$ changes sign twice pointing to possible electronic phase-transitions. e-f) Representative traces of the unconventional THE ($\rho_{xy}^{T,u}$) when the magnetic fields and the electrical currents are oriented within the basal plane. The sign of $\rho_{xy}^{T,u}$ was chosen to be positive. g-h) Contour plots displaying the magnitude of the conventional and unconventional THE responses, respectively. THE is particularly strong at room temperature in both plots.

The conventional Hall coefficient $R_0$ reveals two changes in sign upon cooling (Figure 2d). The first change occurring near 230 K, and probably resulting from the coexistence between electrons and holes with each type of carrier being characterized by distinct temperature dependent mobilities. The second change in sign occurs upon approaching $T_s$ and therefore can be attributed to the magneto-structural transition and its effects on the Fermi surface of this compound.

For a more detailed understanding of the anisotropy in $Fe_{5-x}GeTe_2$, we include the behavior of the u-THE as function of the magnetic field and for several temperatures, in this case from a second crystal that was carefully aligned to have $\mu_0 H$ along a planar direction (Figure 2e). The maxima of the u-THE, i.e., $\rho_{xy}^{T,u,max}$ as a function of $\mu_0 H$ and $T$ (Figure 2f) behaves quite distinctly with respect to the c-THE, $\rho_{xy}^{T,max}(T)$; it remains nearly constant for $T_c \geq T \geq T_s$, decreasing by a factor < 2 below $T_s$. This indicates that the intrinsic anisotropy of this compound affects the phase-diagram and the nature of its field-induced topological spin textures. This assertion is also supported by the contour plots depicting both the c-THE (Figure 2g) and the u-THE (Figure 2f) as functions of magnetic field and temperature. The c-THE displays a sharp maximum in its absolute value that is considerably broader in $\mu_0 H$ relative to the one shown by the u-THE, suggesting again distinct phase diagrams for both orientations. One can see that the behavior of the c-THE is affected by the magneto-structural transition at $T_s$, as is also the case for the u-THE with this being less apparent through its contour plot. This evidence for a c-THE at room temperature lies in direct contrast to previous reports indicating its existence in $Fe_{5-x}GeTe_2$ within a narrower temperature region, i.e., 120 K $\leq T \leq$ 250 K[2a]. Perhaps, the nature of this apparent discrepancy lies in the differences among crystals synthesized through distinct protocols. Data previously reported by Gao *et al*.[2a] seem to have been collected from samples slowly cooled from the growth temperature, or annealed at elevated temperatures, based on their relatively low $T_c$ and the distinct behavior of their magnetic susceptibility as a function of $T$ when compared to our quenched samples[20]. Therefore, the apparent absence of the c-THE at room temperature, as reported by Gao *et al*.[2a] is, in our opinion, related to the synthesis protocol followed to obtain their $Fe_{5-x}GeTe_2$ crystals.

A previous study reported the observation of an u-THE in the sister compound $Fe_{3-x}GeTe_2$, displaying a pronounced peak in the vicinity of ~ 4.5 T[14] for magnetic fields aligned along the electrical currents. Remarkably, this peak is accompanied by concomitant peaks observed in both the Nernst and the thermal Hall response. Thermal transport, in particular the thermal Hall effect $\kappa_{xy}$, is a rather sensitive technique to probe the topological nature of any given compound, given that $\kappa_{xy}$ is directly proportional to the Berry curvature[21] intrinsic to electronic or magnon dispersing bands in conductors or magnetic insulators, respectively. In Figure S6 we provide a summary of our thermal transport study in $Fe_{5-x}GeTe_2$, unveiling that:

i) the anomalous Nernst signal $S_{xy}^A$ in Fe$_{5-x}$GeTe$_2$ exceeds the one extracted for Fe$_{3-x}$GeTe$_2$ leading at $T$ = 80 K to a Nernst angle $\theta_N = \tan^{-1}(S_{xy}^A/|S_{xx}|) = 0.14$ radians that is considerably larger than $\theta_N = 0.09$ reported for Fe$_{3-x}$GeTe$_2$ in Ref. [22], implying a role for both topology and electronic correlations (Figure S7), ii) $S_{xy}^A$ displays a maximum at $T_s$ that is remarkably magnetic field dependent, iii) both the Seebeck $S_{xx}$ and the Nernst $S_{xy}^A$ effects display a $T$-dependence that mimics the electrical transport meaning that they also display three distinct regimes as a function of $T$.

To understand these, and the origin of the THE in Fe$_{5-x}$GeTe$_2$, we performed LTEM at cryogenic temperatures using a liquid nitrogen sample holder, allowing for the collection of data at temperatures as low as 100 K (**Figure 3**). These images were collected from crystals exfoliated (less than 100 nm thick) under inert conditions and encapsulated by a thin graphite top layer to prevent both sample degradation and charge accumulation under the electron beam. In the subsequent discussion, the data collected and described here corresponds to the behavior of the magnetic domains within the basal plane of Fe$_{5-x}$GeTe$_2$. When the sample is cooled from room temperature to 200 K under magnetic field cooled conditions, i.e., under $\mu_0 H = 30$ mT, one observes the progressive emergence of magnetic domains (Figure 3a) which, according to the magnetic induction map (Figure 3b), have essentially a planar component. Notice the presence of domain walls (white or dark lines in Figure 3a) meeting at the boundary between multiple domains and yielding both light and black spots (indicated by magenta and white circles in Figures 3a and 3b, respectively). As we discuss below, these spots are located within the cores of planar magnetic vortices with our micromagnetic simulations shown below pointing to merons. Notice that this observation is consistent with a previous reports revealing the existence of meron and anti meron chains in this compound[2a, 12c]. We are unable to observe the chains given that these are mainly oriented along the interlayer $c$-axis, whereas the images presented here were collected from the basal plane of the material. Before further discussing this point, we present LTEM images collected under field at 100 K (near $T_s$) which, as we argue below, reveal the presence of skyrmions (the light purple region). Coexisting in-plane domains (light green region) are observed when the sample is tilted by -30° (Figure 3c). Note that the skyrmions are only observed in the LTEM images upon tilting the sample, thereby confirming the Néel-type behavior of the domains (Figures S7 and S8). Figure 3d exhibits the in-plane magnetization orientation of the core and surrounding stray fields of Néel skyrmions.

At 100 K, the out-of-plane domains coexist with planar oriented domains (see, region enclosed by the green rectangle) leading to the simultaneous observation of magnetic vortices of distinct structures (Figure 3e). The domain walls and by extension, the magnetic vortices, between planar domains show an increase in contrast at small tilt angles, as shown in Figure 3h for α = -12° in contrast to the relatively poor contrast for the planar domains (Figure 3c). For an illustration on the role of in-plane magnetic fields, or tilt angles on the contrast, allowing us to observe either skyrmions or planar magnetic vortices (in reality, merons), see Figure S8. To illustrate the emergence and coexistence of different types of domains and associated topological spin textures as a function of temperature, we collected Lorentz TEM images under zero field cooled and field cooled conditions (Figures S8 and S9) revealing: i) an increase in contrast among planar domains upon cooling, ii) their weakening near room $T$ due to the application of a modest inter-planar magnetic field for field-cooled data, and iii) the progressive emergence of labyrinthine domains (and Néel skyrmions under field) upon approaching the magneto-structural transition at $T_s$. Although the domain structure of a ferromagnetic compound like $Fe_{5-x}GeTe_2$ is history dependent, it is important to emphasize that the coexistence of different types of domains and associated spin textures upon cooling below ~170 K was reproduced among several samples and through subsequent thermal cycles of the same samples. This observation would explain the decrease in the planar magnetization observed below ~ 170 K[6] (Figure S2), which was subsequently ascribed to the development of ferrimagnetism[11].Therefore, the magneto-structural transition, is preceded by a reorientation of the magneto-crystalline anisotropy towards the *c*-axis that seemingly contributes to the emergence of the labyrinthine domains as well as skyrmions (Figure S9). Perhaps, these domains correspond to fluctuations preceding, and even contributing to the transition at $T_s$. The planar magnetic vortices are found to be rather robust surviving between 100 K and up to 290 K under $\mu_0 H$ = 30 mT. Due to the possible weakening of ferromagnetism in thin exfoliated crystals, it is reasonable to assert that in bulk samples these domains are likely to be present throughout the entire FM region up to ~310 K, and likely beyond, depending on the sample's precise Curie temperature.

To uncover the possible topological character of the spin textures observed via Lorentz TEM in $Fe_{5-x}GeTe_2$, we performed micromagnetic simulations[23] (**Figure 4**). The starting point of the simulations is the use of Voronoi tessellation of the plane of view with Thiessen polygons. Some of these polygons are characterized by a random in-plane magnetic anisotropy to generate a large population of merons, and are depicted by colored polygons (Figure 4a) while others display an out-of-plane anisotropy (white and black polygons). The first-order magneto-crystalline anisotropy constant of $Fe_{5-x}GeTe_2$ was incorporated into the simulations via the measured anisotropy of its magnetization at low fields using the so-called Sucksmith-Thompson method[24] (Figure S10). For the region showing skyrmions, we assumed the existence of an rectangle. Notice the non-coplanar spin texture at the center of the vortices implying that these are in fact meron, anti-meron pairs. j) Schematics of the sample setup illustrating the direction of the magnetic field $\mu_0 H$ and its planar component $\mu_0 H_\parallel$ upon tilting the sample by an angle $\alpha$, which is used to increase the contrast for either merons or skyrmions. To expose skyrmions, we applied fields inferior to 30 mT which is considerably smaller than the value of ~ 150 mT where the maxima in $\rho_{xy}^{T,c,u,max}$ is observed. Therefore, the density and size of the skyrmions extracted from this Figure will not lead to a correct estimate of their contribution (in the order of 1 $\mu\Omega$ cm) to the topological Hall response of $Fe_{5-x}GeTe_2$, which is also influenced by the merons.

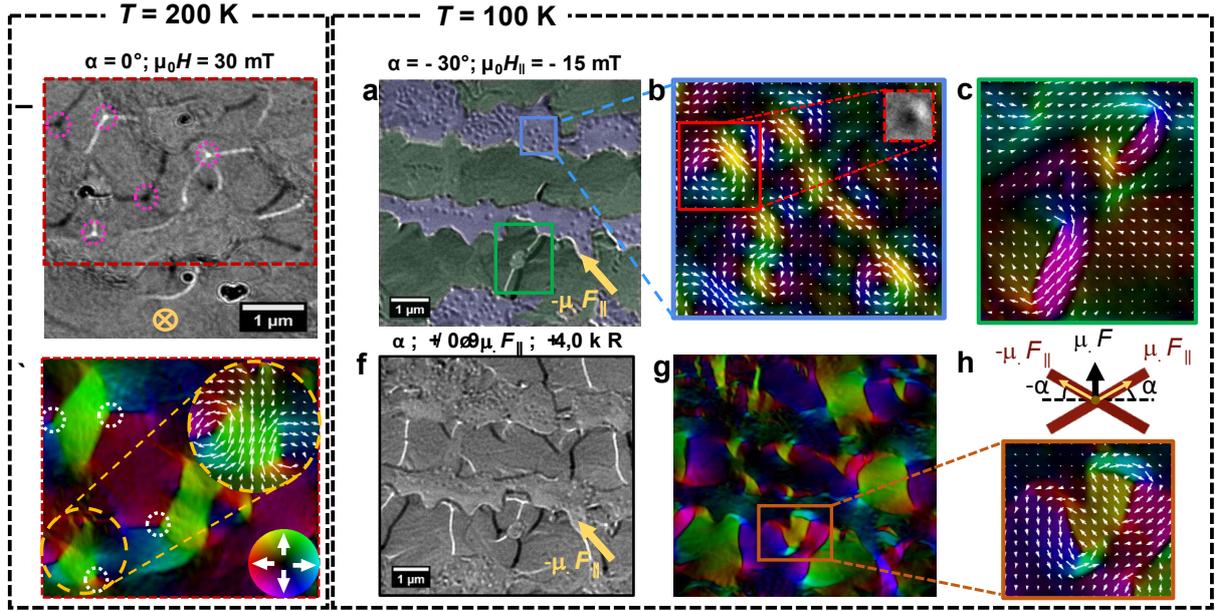

**Figure 3.** Lorentz transmission electron microscopy (LTEM) study of $Fe_{5-x}GeTe_2$. a) LTEM image with a 0° tilting angle collected at $T = 200$ K under $\mu_0 H = 30$ mT, revealing the morphology of planar magnetic domains. Magenta circles encircle white and black dots emerging at the intersection between planar domains. b) Retrieved magnetic induction map of the region enclosed by the red dashed frame in a, indicating the presence of in-plane magnetic domains accompanied by the formation of spin vortices and antivortices at their boundaries (indicated by white circles). White arrows correlated the colors of the induction map with the orientation of the magnetization. Top right inset: magnified induction map of the region enclosed by an orange circle on the bottom left. c) LTEM image from another area from the same sample collected at $T = 100$ K, $\mu_0 H_{\parallel} = -15$ mT, and at a tilt angle $\alpha = -30°$. Different colors are assigned to highlight the coexistence of magnetic domains with in-plane spins (green shaded regions) and domains with out-of-plane moments (purple shaded regions). Faint yellow arrow indicates the direction of the planar component of the magnetic field (see inset in j). d-e) Local view of the magnetic induction maps enclosed by the blue and green rectangles in **c**, respectively. Néel skyrmions are identified as the spin textures contained in d, whereas magnetic vortices are revealed in e. In panel d, the red square to the left encloses an area displaying the orientation of the planar magnetization associated to the core as well as the surrounding stray fields of a Néel skyrmion revealing a typical bound vortex-antivortex structure[25]. Its corresponding LTEM magnetization map is shown in the inset (red square to the right). h) Same area as in c, but under $\mu_0 H_{\parallel} = -6.2$ mT, and $\alpha = -12°$. i) Magnetic induction map of h, showing a mild contrast between the in-plane and the out-of-plane domain configurations. Vortices and antivortices at the domain boundaries are enclosed by the brown

interfacial Dzyaloshinskii–Moriya interaction term of 1.2 mJ/m$^2$ and a uniform micromagnetic exchange constant $A = 1 \times 10^{-11}$ J/m. For a tilt angle $\alpha = 20°$, this magnetic domain structure yields a simulated LTEM contrast (Figure 4b) very similar to the ones seen experimentally (Figure 3, Figures S8 and S9). The same can be said about the reconstructed magnetic induction map (Figure 4c). Most importantly, in these simulations we can study the presence of spin textures with different topological numbers such as those enclosed by blue and red rectangles in Figure 4b[26]. We observed that both skyrmions (Figure 4d) and merons (Figure 4e) are present simultaneously and spread out through the entire surface. The calculation of the topological numbers for these spin textures resulted in magnitudes that are nearly integer ($N \approx \pm 1$) for skyrmions and half-integer ($N \approx \pm 1/2$) for merons (see Supplementary Tables S1-S2). Interestingly, the skyrmions and merons are in areas with out-of-plane and in-plane anisotropies, respectively. This indicates that the coexistence of both spin textures is related with the stabilization of parts of the crystal with different magnetic domain features below $T \approx 160$ K and as $T$ approaches $T_s$. We emphasize that meron chains in $Fe_{5-x}GeTe_2$ have already been reported in Refs. [2a, 12c] through Lorentz TEM by sculpting lamellas and observing their spin textures. Coexistence of merons and skyrmions might result from either local strain associated to the transition, emergence of magnetic domains associated with the low temperature magneto-structural phase, or from local oscillations in the Fe stoichiometry that can stabilize or suppress the previously reported Fe(1) ordering. Preliminary electron energy dispersive spectroscopy (EDS) measurements detect minor fluctuations in the Fe content between the regions displaying distinct types of domains (Figure S14). Fe vacancies would affect the coupling between all three inequivalent Fe sites, and hence favor a distinct local magnetic order. We must emphasize that Néel skyrmions are not observed in our samples at temperatures exceeding $T = 170$ K.

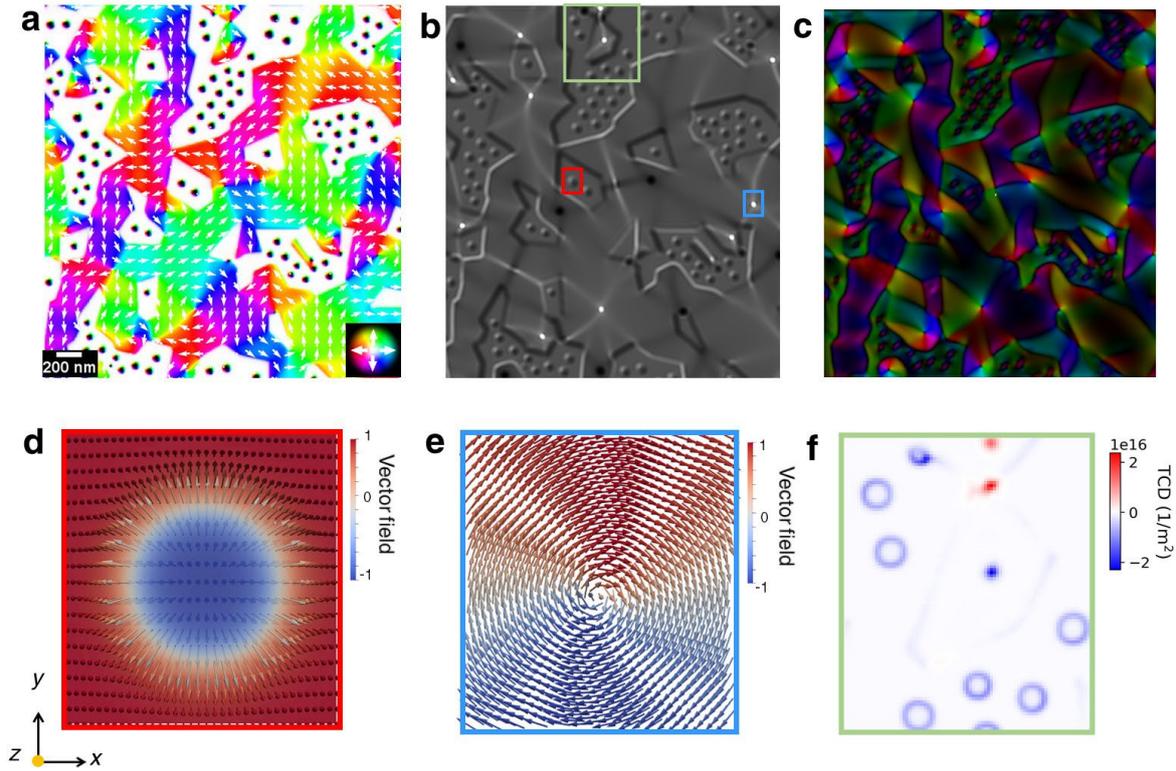

**Figure 4.** Coexistence between merons and skyrmions in $Fe_{5-x}GeTe_2$ according to micromagnetic simulations. a) Magnetization configuration of micromagnetic simulations showing in-plane magnetic domains coexisting with Néel skyrmions (dots). Black and white colors represent out-of-plane magnetization areas pointing inward and outward, respectively. Colored regions depict magnetic domains with in-plane orientation of the spins. b) Simulated Lorentz TEM contrast for the planar domains, merons, and skyrmions in panel a, which was calculated with the sample tilted at $\alpha = 20°$. The Lorentz contrast is sensitive only to circularity of the vortex spin texture (bright dots). c) Reconstructed magnetic induction map from the Lorentz TEM contrast in b. Observed spin textures match those of the in-plane regions in a as well as those in the experimental data (Figure 4i). d) Local spin texture for the enclosed red square in b. The calculated topological number resulted in a value of -0.881 characteristic of a skyrmion-like particle. e) Spin texture enclosed by the blue dot in b. The calculated topological charge yields a value of -0.498, which is consistent of merons. f) Topological charge density for the area enclosed by the green rectangle in b as calculated via the MuMax3 software[23]. Figures S11, S12 and S13 and Tables S1 and S2 provide respectively, additional details on the parameters used for the simulations and resulting analysis[24] (extracted topological charges).

This suggests that their existence in $Fe_{5-x}GeTe_2$ becomes viable only when the samples display domains with predominant out-of-plane spin orientation albeit coexisting with planar domains characterized by in-plane magneto-crystalline anisotropy leading to interfacial DM interaction (Figure 3). Notice, as shown in Figure S2, that the planar component of the magnetization, collected at low fields, decreases sharply below 170 K. This is perfectly consistent with a rotation of the magnetic uniaxial anisotropy towards the *c*-axis and hence with the coexistence of both types of domains below this temperature. A word of caution should be given here since LTEM requires the use of thin samples which might not necessarily display the exact same magnetic phase diagram and spin textures as the bulk single crystals. This is illustrated by Ref.[12b] detecting skyrmions only within a narrow range of thicknesses $t$, i.e., 100 nm $\leq t \leq$ 500 nm in exfoliated Co doped $Fe_{5-x}GeTe_2$. As such we take the LTEM data as guidance for understanding the bulk c-THE and u-THE response. However, we hope our results will stimulate further experimental efforts to elucidate the precise magnetic phase diagram of $Fe_{5-x}GeTe_2$.

As for the observation of skyrmions in a centrosymmetric system like $Fe_{5-x}GeTe_2$, notice that such objects were originally predicted[27] and subsequently observed in centrosymmetric albeit magnetically frustrated systems like, $Gd_2PdSi_3$[28], $Gd_3Ru_4Al_{12}$[29], and $GdRu_2Si_2$[30] and claimed to result from competing, frustrated magnetic interactions. In a previous report[14] on the sister compound $Fe_{3-x}GeTe_2$ we provided Monte Carlo simulations based on a Hamiltonian that included several positive and negative exchange interactions between both inequivalent Fe sites, as well as Dzyaloshinskii-Moriya interaction terms, in addition to biquadratic and uniaxial anisotropy terms. All these interactions compete to stabilize labyrinthine domains as well as field-induced skyrmions in $Fe_{3-x}GeTe_2$. In $Fe_{5-x}GeTe_2$, the presence of three inequivalent Fe sites might add multiple competing interactions among these neighboring Fe sites. Further theoretical and experimental work is needed to clarify the role of competing interactions.

Nevertheless, as we show here, the spin re-orientation transition observed below 160 – 180 K leads to the coexistence of domains having moments predominantly in the planes with domains having moments oriented out of the planes. Such a configuration of domains should locally, and probably also globally, break inversion symmetry favoring the Dzyaloshinskii-Moriya interaction, which in turn favors the stabilization of skyrmions. Therefore, $Fe_{5-x}GeTe_2$ is likely to be characterized by long-range magnetic dipolar, competing exchange interactions, in addition to small DMI terms, all conspiring to stabilize the chiral spin textures observed by us. The role of the dipolar interaction in stabilizing chiral spin textures in $Fe_{5-x}GeTe_2$ could be exposed, for example, via x-ray resonant magnetic scattering which in the case of FePd films revealed magnetic flux closure domains[31], but will be the subject of a future study. However, and in contrast to the other centrosymmetric frustrated magnets, $Fe_{5-x}GeTe_2$ is unique given that it displays a topological Hall response up to room temperature (and beyond), and under rather modest magnetic fields.

## 3. Conclusions

In the context of the topological Hall effect, our findings using an unconventional experimental configuration which, a priori, should not yield any Hall response, meaning magnetic fields and electrical currents along a planar direction, represents a clear advantage with respect to the conventional configuration of measurements. For the unconventional topological Hall-effect configuration of measurements, one does not need to subtract a superimposed anomalous Hall-effect. This implies that no additional analysis over distinct data sets, i.e., transport and magnetization, collected from distinct instruments is necessary. Such manipulation might be prone to instrumental artifacts that should not be present for the unconventional configuration of measurements used in our study. Nevertheless, it is possible, and even likely, that one stabilizes distinct chiral spin textures when the external magnetic field is oriented along the conducting planes of $Fe_{5-x}GeTe_2$ relative to a perpendicular direction. This might explain the differences in the temperature dependence of the conventional and unconventional topological Hall effects as observed through Figures 2g and 2h. A detailed study on the possible chiral spin textures stabilized by fields oriented along a planar direction and leading to the observation of the u-THE will be the subject of future work.

It is worth mentioning that calculations of the stray, or dipolar fields, from the different spin textures (e.g., skyrmions and merons) yielded values within the range of (0.18 – 0.40) T for skyrmions under zero applied field ($H_z$=0 mT), and (0.21 – 0.60) T under $H_z$ = 160 mT (Figure S15). For merons the calculations yielded remarkably large values of (0.10 – 0.40) T under $H_z$ = 0 mT, and (0.3 – 0.7) T under $H_z$=160 mT (Figure S16). These ranges are in remarkable agreement with the largest magnitude displayed by the conventional topological Hall effect $\rho_{xy}^T$ (Figure 2a) and the unconventional THE $\rho_{xy}^{T,u}$ (Figure 2e) response, as captured by the measurements. As the Hall resistance $R_{xy}$ is proportional to the topological number $N_{Sk}$ via $R_{xy} \propto \int_{-y_0}^{y_0} \int_{-x_0}^{x_0} N_{Sk}(x-x', y-y')dx'dy'$ [32], which is also related to the emergent field from the spin textures via $\boldsymbol{B}_{em} \propto N_{Sk}\hat{e}_z$ [33], then the Hall resistance is directly proportional to the emergent field induced by the spin textures. We remark that the experimental Hall signals ($\rho_{xy}^T$, $\rho_{xy}^{T,u}$) in Figures 2a and 2e result from a joint response by skyrmions and merons that cannot be separated. The important point is that our micromagnetic simulations indicate the induction of a large dipolar or emergent field, by both types of spin textures, via the application of a quite modest magnetic field that is strikingly close to the experimental value where one observes the maxima in both $\rho_{xy}^{T}$ and $\rho_{xy}^{T,u}$. An evaluation of the topological Hall response from the emergent fields associated to each type of topological spin texture is provided in the SI file (Figures S17 and S18). This suggests the active role of the topological spin textures on the stabilization of the THE in $Fe_{5-x}GeTe_2$.

We are not aware of any other study that correlates topological transport properties with the observation of merons at room temperature and beyond, or report their coexistence with skyrmions in a particular material that does not involve the stacking of different compounds[34]. This makes the field of 2D vdW magnets fruitful for landmark explorations searching for the stabilization of hybrid spin textures and their possible manipulation via external stimuli such as current and light. Our results suggest that unconventional topological spin textures[10b], that is, those distinct from merons or skyrmions, might exist in atomically thin vdW layers and their properties have yet to be unveiled and explored for spintronic real applications. To support this assertion, we estimated through Lorentz microscopy[35] the domain wall width among planar domains obtaining a remarkably wide average width df = (25 ± 5) nm (Figure S19). As such a wide domain wall meanders between planar domains, it is likely to locally acquire either Néel or Bloch character likely explaining the apparent discrepancies among the different reports on $Fe_{5-x}GeTe_2$[36]. This hybrid character would be susceptible to the application of an external magnetic field and contribute to the novel topological transport observed in the $Fe_{n-x}GeTe_2$ compounds.

In $Fe_{5-x}GeTe_2$, the strong coupling between the electronic and thermal transport properties to topological spin textures that are pervasive over a wide range of temperatures makes this system a promising candidate for applications in skyrmionics and may lead to a new field, that of "meronics". A topological Hall-effect at and beyond room temperature coinciding with the existence of topological spin textures may provide opportunities for the field of skyrmionics based on 2D materials[37]. To this regard, $Fe_{5-x}GeTe_2$ and particularly its doped variants[12b, 38] emerge as serious candidates for the possible development of applications in spintronics, given that they can be grown in large area[39], display Curie temperatures exceeding room temperature[38], and display crystal thickness[40] dependent skyrmions sizes.

## 4. Experimental Section

*Single-crystal synthesis*: Single crystals of $Fe_{5-x}GeTe_2$ were synthesized through a chemical vapor transport technique. Starting molar rations of 6.2:1:2 for Fe, Ge, and Te respectively, were loaded into an evacuated quartz ampoule with approximately 100 mg of $I_2$ to act as the transport agent. After the initial warming, a temperature gradient of 75°C was established between a 775 °C and 700 °C zone of a 2-zone furnace and maintained for 14 days, during which large single crystals nucleated at the 700°C zone. Samples were subsequently quenched in ice water to yield the maximum Curie temperature[6]. Crystals used in this paper are from the same batch used in previous experiments[7]. Crystals were washed in acetone and subsequently isopropyl alcohol to remove residual iodine from their surface. According to Energy Dispersive Spectroscopy the values of *x* are found to oscillate between 0.15 and 0.

*Electrical transport measurements:* Platinum wires with a diameter of 25 μm were fixed onto deposited gold pads via silver paint. The Au pads were deposited via magnetron sputtering on freshly cleaved surfaces of $Fe_{5-x}GeTe_2$ to minimize the effects of oxidation or residual iodine on the as grown surface. To prevent oxidation, single crystals were exfoliated under argon atmosphere, within a glove box containing less than 10 parts per billion in oxygen, and water vapor. These were subsequently dry transferred onto gold on chromium contacts pre-patterned on a $SiO_2$/*p*-Si wafer using a polydimethylsiloxane stamp, and subsequently encapsulated among *h*-BN layers, with both operations performed under inert conditions. Chromium and gold layers were deposited via e-beam evaporation techniques, and electrical contacts fabricated through electron beam lithography. All measurements were performed in a Quantum Design Physical Property Measurement System.

*Thermal transport measurements*: Thermal conductivity and the thermal Hall effect were measured using a one-heater three-thermometer method. Additional electrical contacts allowed us to measure four-probe resistivity, Hall effect, Seebeck, and Nernst effects simultaneously. For the thermal transport measurements, a heat pulse was applied to generate a longitudinal thermal gradient corresponding to a 3% of the sample base temperature. After applying the heat pulse, the temperature of all three thermometers were monitored until they reached a stable condition (defined as a rate of less than 1 μK/s) averaged over 15s. Typical timescales were 5 to10 s for temperature rise and 30 to 60 s for its stabilization. A step wise increase in heat was also applied to generate corresponding stepwise thermal gradients, from which a linear relation between the measured values (e.g., thermal electromotive force as a function of temperature gradients for Seebeck and Nernst effects; temperature gradients as a function of heat power) was used to obtain the relevant thermal transport variables. The results from both methods are practically identical. The measurements were performed in a Quantum Design physical properties measurement system (Quantum Design PPMS), which allowed in situ calibration of thermometers in the presence of exchange gas followed by thermal measurements under high vacuum.

*Cryogenic Lorentz transmission electron microscopy*: Single crystalline $Fe_{5-x}GeTe_2$ was mechanically exfoliated directly onto homemade polydimethylsiloxane stamp inside an argon filled glovebox. Prior to its utilization, the stamp was rinsed in acetone and isopropyl alcohol to clean its surface. After appropriate crystal thicknesses and dimensions were identified via optical contrast, the selected crystal(s) was transferred onto a window of a silicon-nitride based transmission electron microscopy grid. Few-layer graphite (14 nm thick) was transferred onto the $Fe_{5-x}GeTe_2$ flake through the same dry transfer method to protect the sample from oxidization. To characterize the magnetic domains, the out-of-focus LTEM images were taken on an aberration corrected JEOL ALTEM2100F Lorentz TEM, which is free of magnetic field at the sample position and is used for the ZFC experiment. Field cooling measurement was carried out in a JEOL 2100F TEM operating in Lorentz mode (Low Mag), a perpendicular magnetic field aligned parallel to electron beams being generated by applying a small amount of current to the objective lens. The magnetic induction maps were reconstructed based on transport-of equation (TIE ) method using the PyLorentz software package[25].

*Statistical Analysis:* No statistical analysis was applied to the data collected and displayed throughout this manuscript. The topological Hall response of $Fe_{5-x}GeTe_2$ was observed in 6 distinct single-crystals, thus confirming its magnitude and reproducibility. It was also measured in 6 exfoliated and encapsulated crystals. The magnitude of the topological Hall response was found to be sample thickness dependent for samples having thicknesses inferior to ~70 nm.

**Supporting Information**

Supporting Information is available from the Wiley Online Library or from the corresponding author.

**Acknowledgements**

L.B. acknowledges support from the US DoE, BES program through award DE-SC0002613 US (synthesis and measurements), US-NSF-DMR 2219003 (heterostructure fabrication) and the Office Naval Research DURIP Grant 11997003 (stacking under inert conditions). The National High Magnetic Field Laboratory acknowledges support from the US-NSF Cooperative agreement Grant numbers DMR-1644779 and DMR-2128556, and the state of Florida. EJGS acknowledges computational resources through CIRRUS Tier-2 HPC Service (ec131 Cirrus Project) at EPCC (http://www.cirrus.ac.uk) funded by the University of Edinburgh and EPSRC (EP/P020267/1); ARCHER UK National Supercomputing Service (http://www.archer.ac.uk) via Project d429. EJGS acknowledges the Spanish Ministry of Science's grant program "Europa-Excelencia" under grant number EUR2020-112238, and the EPSRC Open Fellowship (EP/T021578/1). Work at Argonne was funded by the US Department of Energy, Office of Science, Office of Basic Energy Sciences, Materials Science and Engineering Division. Use of the Center for Nanoscale Materials, an Office of Science user facility, was supported by the U.S. Department of Energy, Office of Science, Office of Basic Energy Sciences, under Contract No. DE-AC02-06CH11357.

**Conflict of Interest**

The authors declare no conflict of interest.

**Data availability**

The datasets generated during and/or analyzed during this study are available from the corresponding author upon request.

TOC

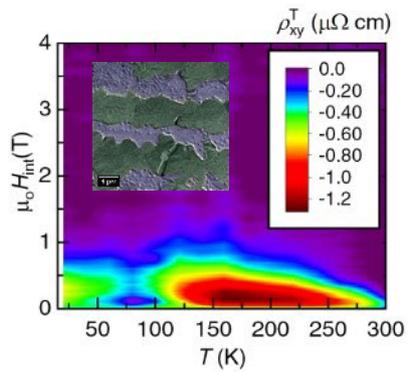

TOC Figure. Contour plot displaying the magnitude of the conventional topological Hall effect revealing that it is particularly strong up to room temperature. Inset: Lorentz-TEM image from an area of the sample collected at $T = 100$ K, highlighting the coexistence of magnetic domains having in-plane oriented spins (green shaded regions) with out-of-plane ones (purple shaded regions).

# Supporting Information for "Coexistence of merons with skyrmions in the centrosymmetric van der Waals ferromagnet Fe$_{5-x}$GeTe$_2$"

Brian W. Casas[1,†], Yue Li[2,†], Alex Moon[1,3], Yan Xin[1], Conor McKeever[4], Juan Macy[1,3], Amanda K. Petford-Long[2,5], Charudatta M. Phatak[2,5], Elton J. G. Santos[4,6], Eun Sang Choi[1], and Luis Balicas[1,3,*]

[1]National High Magnetic Field Laboratory, Tallahassee, 32310, FL, USA.
[2]Materials Science Division, Argonne National Laboratory, Lemont, 60439, IL, USA.
[3]Department of Physics, Florida State university, Tallahassee, 32310, FL, USA.
[4]Institute for Condensed Matter and Complex Systems, School of Physics and Astronomy, The University of Edinburgh, EH9 3FD, UK.
[5]Department of Materials Science and Engineering, Northwestern University, Evanston, 60208, IL, USA.
[6]Higgs Centre for Theoretical Physics, The University of Edinburgh, EH9 3FD, UK.

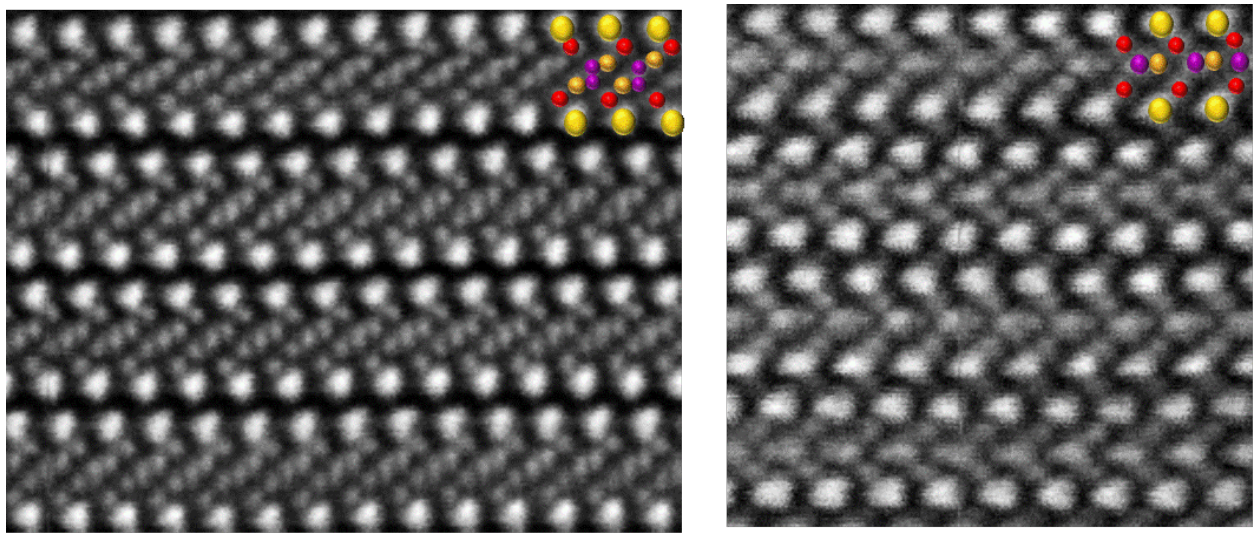

**Figure S1**. Atomic resolution HAADF-STEM image of Fe$_{n-x}$GeTe$_2$ along [100] direction. Left: High angle annular dark field scanning transmission electron microscopy (HAADF-STEM) image corresponding to a transversal cut of a Fe$_{5-x}$GeTe$_2$ single crystal, showing the conducting/magnetic planes separated by van der Waals gaps. Right: HAADF-STEM image of the transversal section of a Fe$_{3-x}$GeTe$_2$ single crystal. Here, yellow, purple, red, and orange dots depict Te, Ge, Fe(1), and Fe(2 as well 3) atoms. The Fe$_{3-x}$GeTe$_2$ image is provided to allow a comparison with the one collected from Fe$_{5-x}$GeTe$_2$.

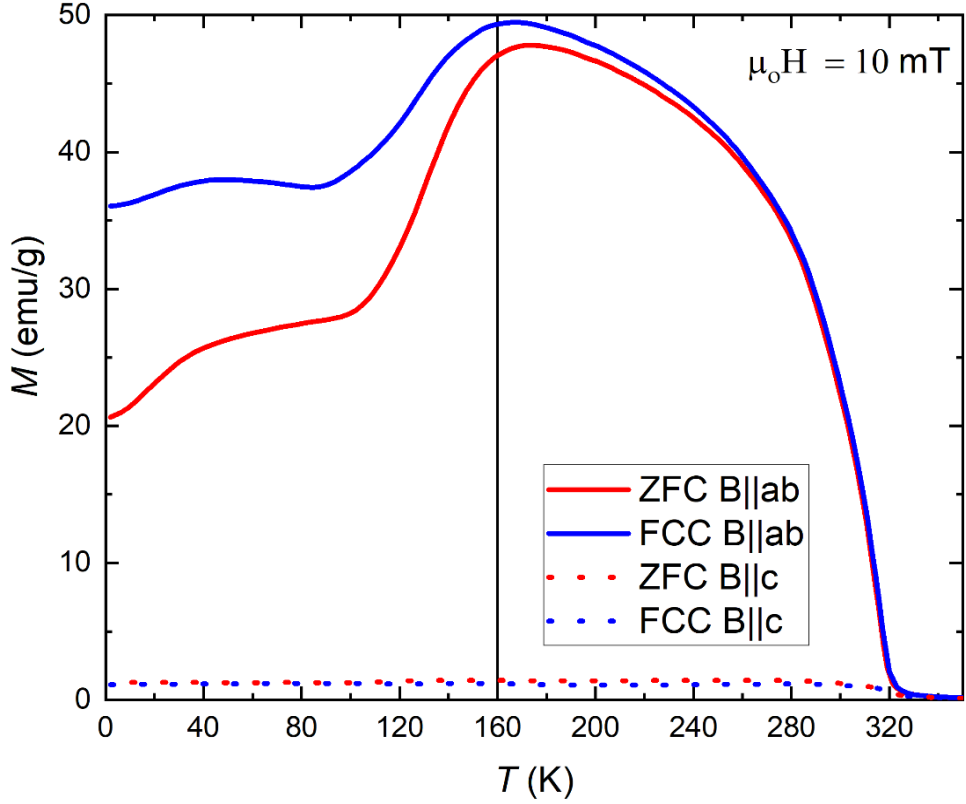

**Figure S2.** Evolution of magnetization as a function of the temperature in single crystalline $Fe_{5-x}GeTe_2$. Both ZFC and FCC protocol are used for orientations of field in-plane (solid lines) and out-of-the-sample plane (dashed lines).

1. Extraction of the conventional topological Hall effect

To calculate the conventional topological Hall effect requires anti-symmetrization of the raw Hall data, to remove any superimposed longitudinal magnetoresistivity signal. Subsequently, one must evaluate and subtract the normal Hall $\rho_{xy}^N$ to obtain the anomalous Hall signal $\rho_{xy}^A$ which contains a superimposed topological Hall component due to spin chirality. Prescriptions for this procedure
have been previously described elsewhere and is described below through Figure S3.

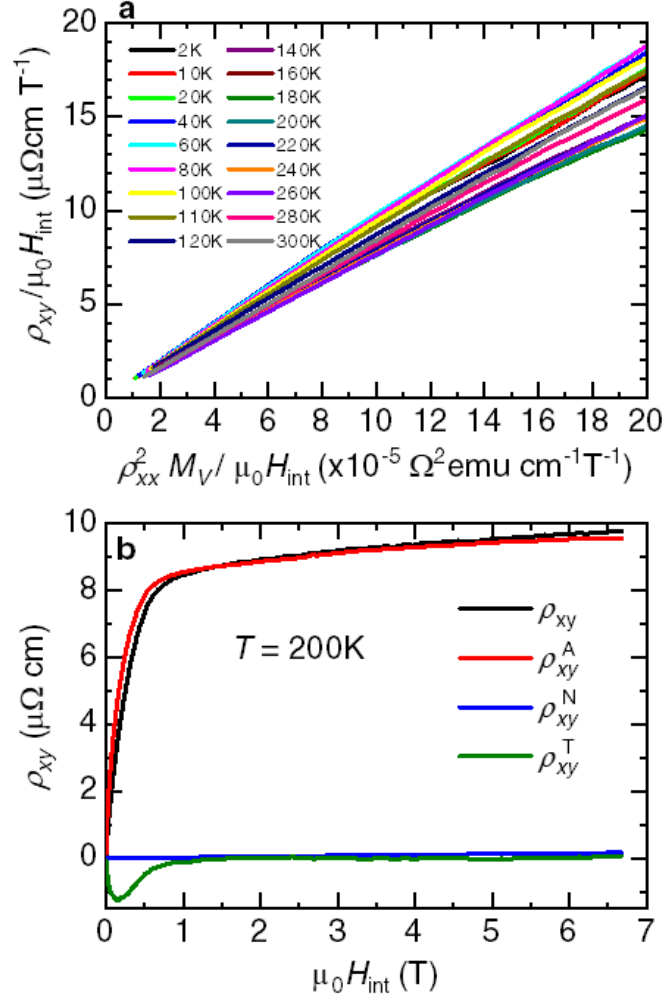

**Figure S3.** Extraction of the conventional topological Hall-effect. a) $\rho_{xy}$ as a function of $\rho_{xx}^2 M_v$ with both axes normalized by the effective magnetic field $\mu_0 H_{int}$ defined by the sample's demagnetization factor. The observed linear dependence yields both $R_0$ and $S_H$. b) Once $S_H$ is known, one can rescale $\rho_{xx}^2 M_V(\mu_0 H)$ by $S_H$ to match the saturation value of $\rho_{xy}^A(\mu_0 H)$. Their difference yields the conventional topological Hall response $\rho_{xy}^T(\mu_0 H)$. In this panel, black trace corresponds to the raw Hall signal $\rho_{xy}(\mu_0 H)$, red trace to $\rho_{xy}^A(\mu_0 H)$, blue to the extracted normal Hall response $\rho_{xy}^N(\mu_0 H)$ and green to $\rho_{xy}^T(\mu_0 H)$. Modeling the anomalous Hall is largely the most sensitive part of this process, though it has been shown that to a good approximation $\rho_{xy}^A$ can be well modeled by $\rho_{xx}^2 M_V$, where $\rho_{xx}$ is the isothermal magnetoresistance, and $M_V$ is the volume magnetization in emu/cm³. At sufficiently high magnetic fields, or fields under which the sample is well into the fully spin polarized regime, $\rho_{xy}/\mu_0 H_{int} \sim R_o + S_H(\rho_{xx}^2 M_V/\mu_0 H_{int})$, where $\mu_0 H_{int}$ represents the effective magnetic field sensed by the sample due to its demagnetization factor defined by its geometry. For this work, the demagnetization factor was approximated to a value of ~ 0.77. In effect, for effective magnetic fields $\mu_0 H_{int} \geq 2$ T one observes a well-defined linear behavior of $\rho_{xy}$ on $(\rho_{xx}^2 M_V/\mu_0 H_{eff})$ and as such were used in the extraction of the conventional and anomalous Hall coefficients $R_0$ and $S_H$ respectively.

The extraction of the conventional THE, or c-THE, requires the anti-symmetrization of the raw Hall data to remove any superimposed residual contribution from the longitudinal isothermal magnetoresistivity $\rho_{xx}(\mu_0 H, T = \text{constant})$, to the normal $\rho_{xy}^N$, and anomalous Hall $\rho_{xy}^A$, components. Prescriptions for this procedure have been previously provided elsewhere but modeling the anomalous Hall response turns out to be the most delicate part of this process. Nevertheless, it has been shown that $\rho_{xy}^A$ can be modeled by $\rho_{xx}^2 M_v$[19] to a good approximation, where $M_v$ is the volume magnetization in emu/cm$^3$. At sufficiently high magnetic fields, or when the sample is well into a fully spin polarized regime, $\rho_{xy}^A/\mu_0 H_{int} = R_0 + S_H(M_v \rho_{xx}^2/\mu_0 H_{int})$, where $\mu_0 H_{int}$ represents the effective magnetic field seen by the sample due to the demagnetization factor associated with its geometrical factors. For fields along the interlayer $c$-axis, the demagnetization factor is approximated to 0.77 for the sample studied in Figure 1 in the main text. Effective magnetic fields exceeding $\mu_0 H_{int}$ = 2 T result into a well-defined linear dependence for $\rho_{xy}^A$ on $M_v \rho_{xx}^2/\mu_0 H_{int}$, or precise values for $S_H$, which were used for the extraction of both $\rho_{xy}^N$ and $\rho_{xy}^A$ (Figure S3). The THE component, or $\rho_{xy}^T$, is obtained via a subtraction: $\rho_{xy}^T = \rho_{xy} - \rho_{xy}^A - \rho_{xy}^N$.

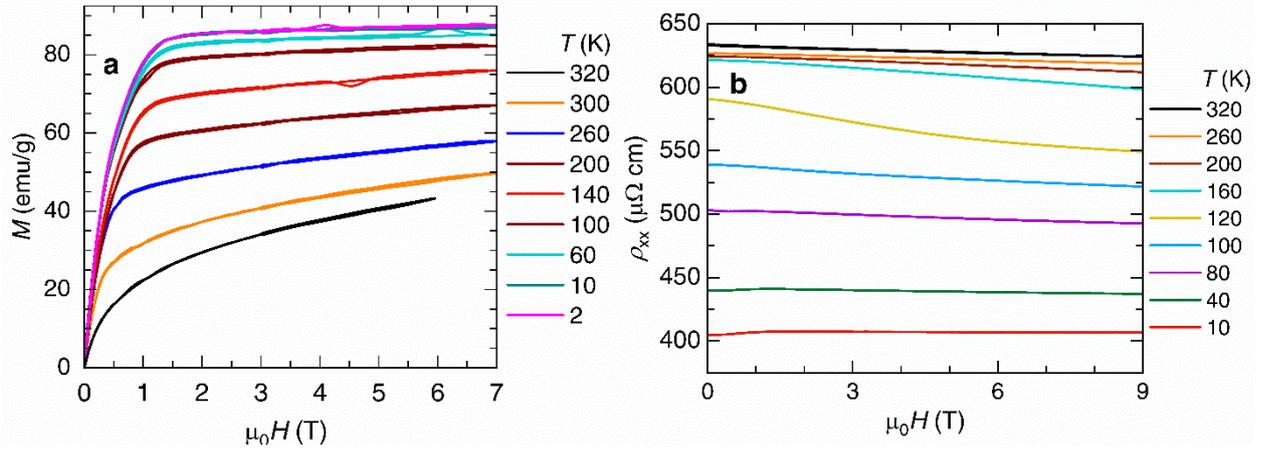

**Figure S4.** Magnetization and magnetoresistivity as a function of magnetic field and temperature for a Fe$_{5-x}$GeTe$_2$ single-crystal. a) Representative magnetization $M$ as a function of magnetic field $\mu_0 H$ traces for the Fe$_{5-x}$GeTe$_2$ single-crystal whose THE data is shown in Figure S1 at several temperatures $T$. b) Magnetoresistivity as function of $\mu_0 H$, again for the same single crystal, and for several $T$s. Here, the field is applied along the interlayer $c$-axis.

2. Unconventional Topological Hall response for fields along the planar direction in an encapsulated 50 nm thick Fe$_{5-x}$GeTe$_2$

To improve the signal to noise ratio, we exfoliated and encapsulated a Fe$_{5-x}$GeTe$_2$ single-crystal under inert conditions. This crystal was encapsulated with a top $h$-BN layer after being transferred on pre-patterned Ti:Au contacts deposited onto a SiO$_2$/$p$-Si wafer. Through atomic force microscopy, this crystal was found to be nearly 50 nm thick. Great care was taken to precisely orient this crystal through a stepper-motor controlled rotator, along a planar direction, via the minimization of the Hall like voltage observed at high magnetic fields.

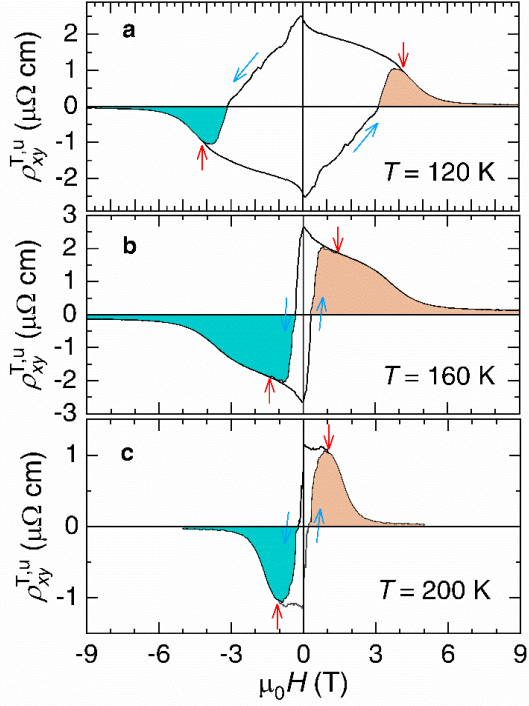 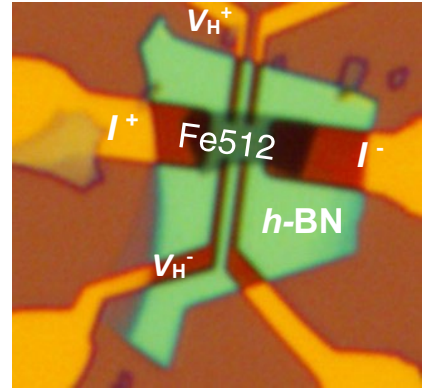

**Figure S5.** Unconventional topological Hall response from an encapsulated $Fe_{5-x}GeTe_2$ crystal. a), b), and c), Hall like response $\rho_{xy}^{T,u}$ observed for fields aligned along the electrical currents in a 15 nm thick encapsulated $Fe_{5-x}GeTe_2$ crystal at temperatures of 120 K, 160 K and 200 K, respectively. Notice the observation of a sizeable, temperature dependent unconventional topological Hall signal observed (areas shaded in green and beige) beyond the coercive field indicated by the red vertical arrows. d) Microphotograph of the 15 nm thick, $Fe_{5-x}GeTe_2$ crystal encapsulated with a top $h$-BN layer, used for these measurements.

As seen in Figure S5, exfoliation leads to the emergence of a large, temperature dependent, coercive field. The recovery of a reversible region as the magnetic field is swept, is indicated by red arrows, while blue arrows indicate the direction of the field sweep. The pronounced irreversibility implies the evolution of the spin textures and an increase in the hardness of the associated magnetic domains upon exfoliation. This would explain the broad peak in $\rho_{xy}^{T,u}$ in the reversible region observed at relatively high fields, since this peak contrasts with the sharper peak observed in $\rho_{xy}^{T,u}$ at very low fields in bulk samples (see main text). Therefore, the origin of $\rho_{xy}^{T,u}$ would be intrinsically associated to spin textures among magnetic domains.

## 3. Anomalous Thermal transport

A previous study reported the observation of an u-THE in the sister compound $Fe_{3-x}GeTe_2$, which displays a pronounced peak in the vicinity of ~ 4.5 T[14] for magnetic fields aligned along the electrical currents. Remarkably, this peak is accompanied by concomitant peaks observed in both the Nernst, $S_{xy} = E_y/(\mu_0 H_z \nabla T_x)$ (where in a conventional configuration $\mu_0 H_z$ would correspond to a field applied perpendicularly to the external temperature gradient $\nabla T_x$) and the thermal Hall, $\kappa_{xy} = j_{Qx}/\nabla T_y$ (where $j_{Qx}$ is a thermal gradient applied along a planar direction and $\nabla T_y$ the gradient in temperature measured along the transverse planar direction) effects in $Fe_{3-x}GeTe_2$.

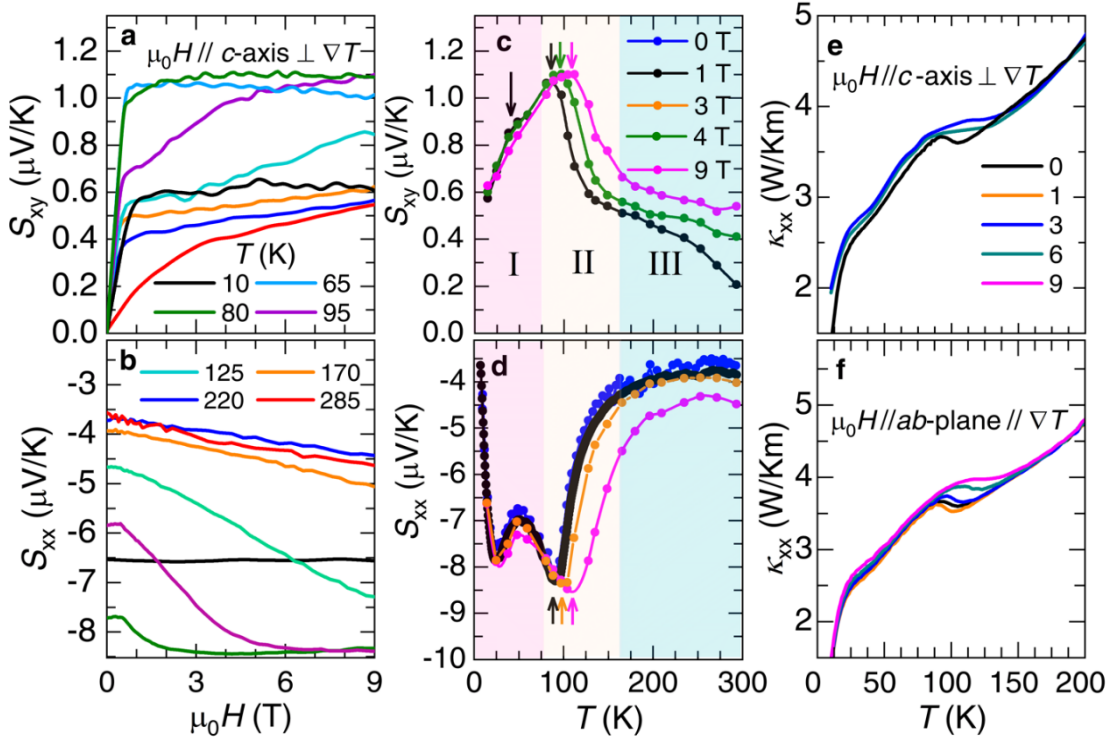

**Figure S6**. Thermal transport in single-crystalline $Fe_{5-x}GeTe_2$. a) Nernst effect $S_{xy}$ as a function of the magnetic field $\mu_0 H$ applied along its $c$-axis. The thermal gradient $\nabla T_x$ is applied in-plane. b) Seebeck coefficient $S_{xx}$ measured simultaneously on the same crystal as a function of $\mu_0 H$. $S_{xx}$ becomes nearly independent of $\mu_0 H$ for $T < T_s \cong 110$ K. $T_s$ corresponds to the magneto-structural transition temperature. c-d) $S_{xy}$ and $S_{xx}$ as functions of $T$ for several values of $\mu_0 H$, respectively. Both quantities display clear anomalies at $T_s$ (indicated by vertical arrows) as well as at $T \cong 25$ K suggesting an additional spin reconfiguration upon cooling. Notice that a magnetic field of 9 T is enough to shift the anomaly up to $T_s$. The color shaded areas highlight regions characterized by changes in the thermal transport that correlate well with the coexistence between skyrmions and merons (region-II), and the exclusive presence of merons (region III) according LTEM. Region-I reveals a reduced Nernst response, implying that the transition at $T_s$ affects the chiral spin textures and therefore the electronic Berry phase of the heat carriers. e-f) Thermal conductivity $\kappa_{xx}$ as a function of $T$ for fields along the $c$-axis and the $ab$-plane, respectively. The anomaly in $\kappa_{xx}$ at $T_s$ is also observed to increase in $T$ as $\mu_0 H$ increases. In e, the temperature dependence of $\kappa_{xx}$ under field was extracted from multiple $\kappa_{xx}(\mu_0 H)$ traces collected under isothermal conditions.

Thermal transport, in particular the thermal Hall effect, is a sensitive technique to probe the topological nature of any given compound, given that $\kappa_{xy}$ is directly proportional to the Berry curvature[21] intrinsic to electronic or magnon dispersing bands in conductors or magnetic insulators, respectively. In the case of $Fe_{3-x}GeTe_2$, $S_{xy}$ was found to change sign upon cooling, which also leads to a change in the sign of $\alpha_{xy}$, the off diagonal component of the thermoelectric conductivity tensor[14] which is directly proportional to the Berry curvature $\Omega_z$. Given the absence of a thermodynamic phase transition observable, for example, through the heat capacity that would explain the change in the sign of $\alpha_{xy}$, we proposed that this observation would result from a topological transition resulting from a reconfiguration of the topological spin textures and their effect on $\Omega_z$. Given the magneto-structural transition at $T_s$ and its effects on the THE (Figure 2 in the main text), it is pertinent to ask if it might affect the topological spin textures and associated thermal transport. To address this point, we performed Nernst, thermal Hall, Seebeck ($S_{xx} = E_x/\nabla T_x$), and thermal conductivity ($\kappa_{xx} = j_{Qx}/\nabla T_x$) measurements in single-crystals of Fe$_{5-x}$GeTe$_2$, as functions of both field and temperature (Figure S6).

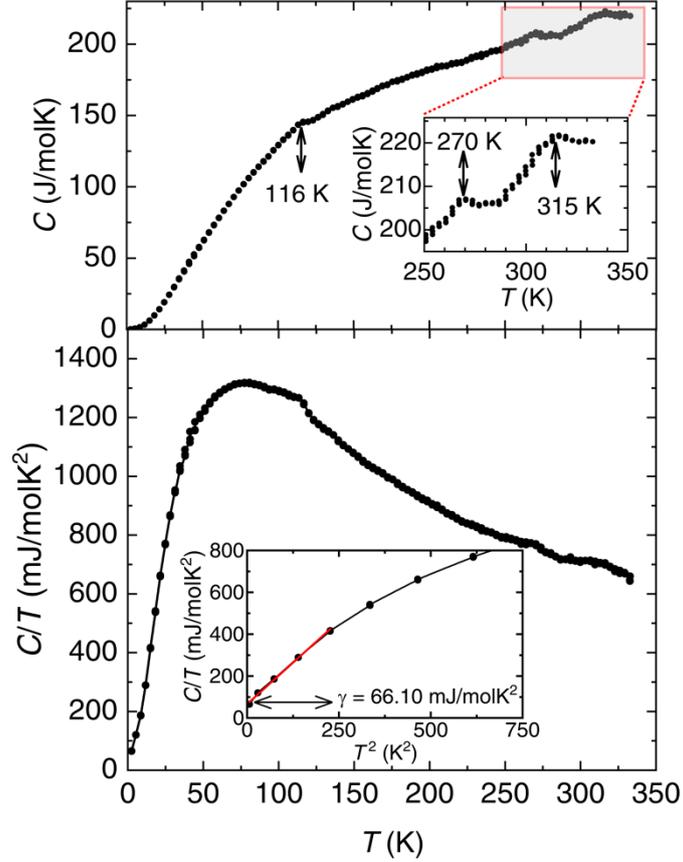

**Figure S7.** Heat capacity $C$ as function of the temperature $T$ for a Fe$_{5-x}$GeTe$_2$ single-crystal. Inset: highlighted region revealing two weak anomalies, one at the Curie temperature at $T_c = 315$ K and a second one suggesting spin reorientation at $T = 270$ K. Bottom panel: C normalized by $T$ as a function of $T^2$. Red line yields the intercept γ corresponding to the electronic contribution to the heat capacity. A coefficient γ = 66 mJ/molK$^2$ points to the relevance of electronic correlations.

Although the anomalous thermal Hall-effect, which is directly proportional to the magnetization, is comparable in size (not shown) to values extracted from Fe$_{3-x}$GeTe$_2$, in Fe$_{5-x}$GeTe$_2$ for 50 K ≤ $T$ ≤ 150 K the saturation value of its anomalous Nernst-effect, $S_{xy}$, is nearly three times larger (Figure S6a). This leads to an anomalous Nernst angle $\theta_N = \tan^{-1}(S_{xy}^A/|S_{xx}|) = 0.14$ radians at $T = 80$ K which is considerably larger than the value 0.09 reported in Ref. [22] for Fe$_{3-x}$GeTe$_2$ that was proposed to result from a large contribution of the Berry curvature near the Fermi level to its anomalous transport variables. To put this value in perspective, the trace collected at $T = 65$ K (Figure S6a) yields a Nernst coefficient $\nu = 1.62$ µV/KT as soon as the field saturates the Nernst response. This value is larger, or comparable, to those collected at lower $T$s for heavy Fermion compounds like CeCoIn$_5$ which were claimed to be "gigantic" relative to those of conventional compounds[41]. This implies that Fe$_{5-x}$GeTe$_2$ displays pronounced electronic correlations, as indicated by its relatively large electronic contribution to the heat capacity $\gamma \cong 66$ mJ/molK$^2$ (Figure S7), which is compounded by the contribution of the topological spin textures on the Berry phase of its charge carriers.

Most likely, both correlations and topological spin textures cooperate to yield a very pronounced Nernst coefficient in Fe$_{5-x}$GeTe$_2$. $S_{xx}$ is negative indicating electron dominated transport, with its magnitude increasing with the external field, thus indicating transport dominated by spin scattering that is suppressed as $\mu_0 H$ increases or as $T$ is lowered below $T_s$ (Figure S6b). In contrast to Fe$_{3-x}$GeTe$_2$, $S_{xy}^A$ is not observed to change its sign (e.g., upon cooling below $T_s$) although it does display a maximum at $T_s$ that is remarkably magnetic field dependent, increasing by ~ 20 K upon application of $\mu_0 H = 9$ T (as indicated by vertical arrows in Figures S6c and S6d).

The field-induced increase in $T_s$ would be compatible with a ferromagnetic state below $T_s$, but less likely to reconcile with a ferrimagnetic state as proposed by Ref.[11]. A second anomaly of unknown origin, likely resulting from a reconfiguration among spin textures, is observed at 25 K, which is more clearly visible in $S_{xx}(T)$ (Figure S6d). These anomalies, and their magnetic field dependence, are quite apparent in both $S_{xx}(T)$ and $\kappa_{xx}(T)$ (Figures S6e and S6f), and seemingly independent of the orientation of $\mu_0 H$ despite the layered nature of this compound. We attempted, unsuccessfully, to detect the effect of topological spin textures on $S_{xy}$, through the observation of a peak concomitant with the one observed in the u-THE, for $\mu_0 H$ aligned along $\nabla T_x$. Its non-observation could have been masked by the relatively large superimposed anomalous Nernst component due to a small sample misalignment.

## 4. Lorentz-TEM

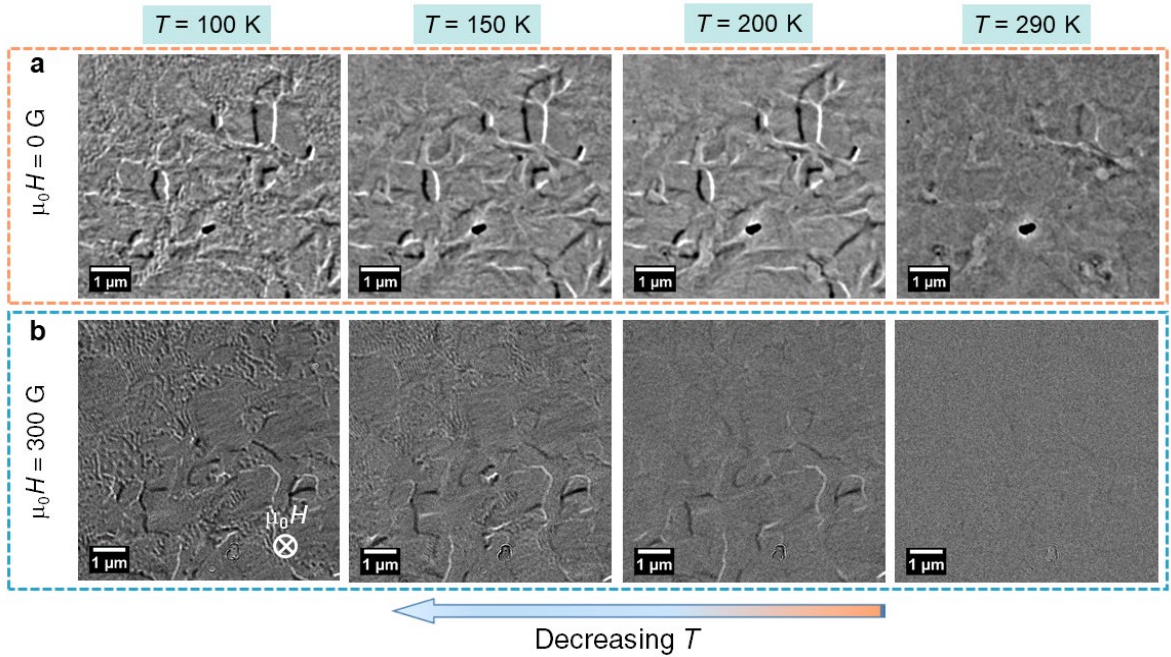

**Figure S8.** Evolution of the in-plane and out-of-plane magnetic domains as a function of temperature and field according to L-TEM. a) In-situ LTEM images as $T$ is lowered following a zero magnetic-field cooling protocol, with a sample tilt angle $\alpha = 20°$ with respect to the horizontal direction. b) LTEM images collected through a field-cooled protocol under a magnetic field of 30 mT, with the sample tilted by $\alpha = -20°$. The defocus length is $\sim -3$ mm. At high $T$s the system exhibits planar magnetic domains that can be suppressed by applying a field of 30 mT. Upon cooling, the contrast due to the spin textures inherent to planar magnetic domains become more pronounced, with labyrinthine domains (under $\mu_0 H = 0$ T) and labyrinthine domains mixed with skyrmions (under $\mu_0 H = 30$ mT) emerging as one approaches the magnetostructural transition at $T_s \sim 110$ K.

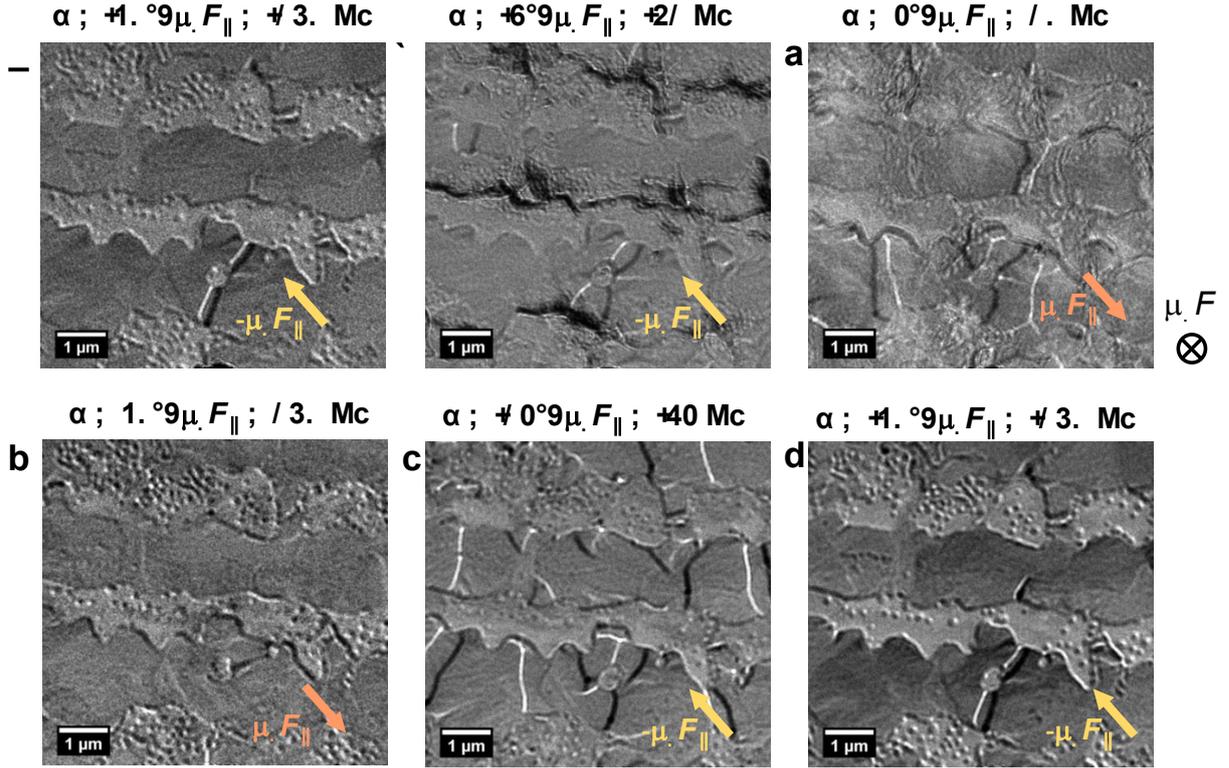

**Figure S9**. Creation of merons and (anti)merons through the manipulation of the in-plane magnetic field ($\mu_0 H_\parallel$). $\mu_0 H_\parallel$ is introduced and controlled by tilting the sample by an angle $\alpha$ under a fixed perpendicular magnetic field $\mu_0 H$ which is aligned along the electron beam. a-f) Lorentz TEM images collected at a temperature $T = 100$ K for several tilt angles $\alpha$ after following a field cooled protocol under $\mu_0 H = 30$ mT g) To image skyrmions, the sample must be tilted. Under these conditions, the in-plane magnetic domains become polarized through the introduced of an in-plane magnetic field that suppresses merons. c) Tilting the sample back to a lower angle of $\alpha = 2°$ drives the formation of multiple planar domains containing merons and (anti)merons at their boundaries. Under these conditions the resulting contrast precludes the observation of skyrmions. This suggests that the observed skyrmions are of Néel type. By sweeping the in-plane field from +15 mT to -6.2 mT (panels d), and e)), one can create a lot of merons and (anti)merons, coexisting with skyrmions.

5. Evaluation of the magneto-crystalline anisotropy used for the micromagnetic simulations

Figure S10 a) shows magnetization $M$ hysteresis loops for Fe$_{5-x}$GeTe$_2$ and magnetic fields along the *ab*-plane and *c*-axis at $T = 80$ K, indicating that the easy axis tends to align along the *ab*-plane at this temperature. The so-called Sucksmith-Thompson method (Ref.[25] in the main text) is used to extract the value of the magneto-crystalline anisotropy. The Sucksmith-Thompson related to first- and second-order magneto-crystalline anisotropy terms for Fe$_{5-x}$GeTe$_2$ system can be expressed as follows:

$$\frac{H}{M_\perp} = \frac{2K_1}{M_s^2} + \frac{4K_1}{M_s^4} M_\perp^2 \qquad (1)$$

where $M_\perp$ is the magnetization perpendicular to the easy axis, $M_s$ is the saturation magnetization along the easy axis, while $K_1$ and $K_2$ are the first- and second-order magneto-crystalline anisotropy constants, respectively. By fitting the measured data to Eq.1, we can estimate the values of both $K_1$ and $K_2$ from the slop and intercept, respectively. The extracted values for $K_1$ and $K_2$ are approximately 2.5 x $10^5$ J/m$^3$ and 9.7 x $10^4$ J/m$^3$. In the micromagnetic simulations, we only consider the contribution of the first-order anisotropy term to the magnetic energy.

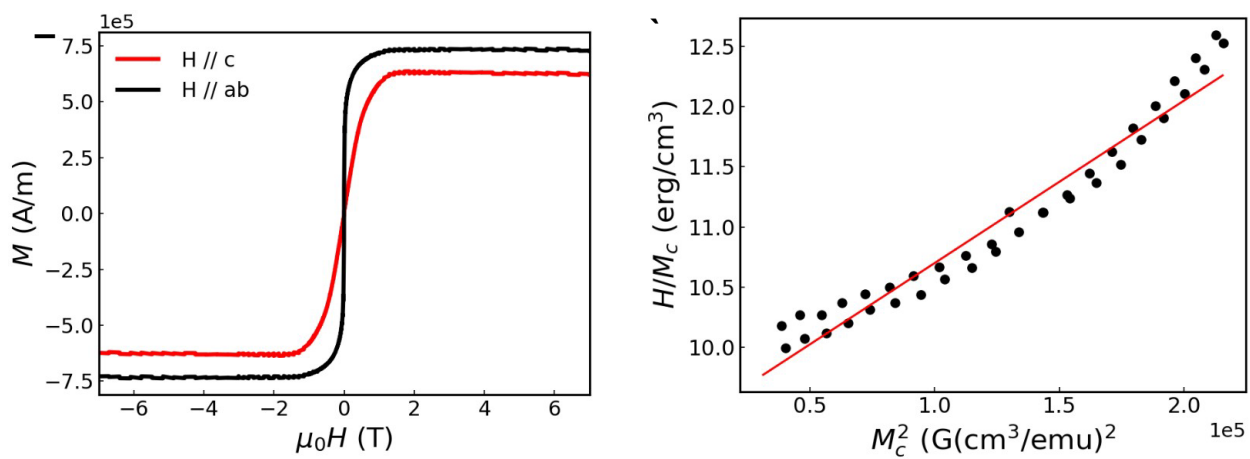

**Figure S10.** Determination of magnetic anisotropy parameters subsequently used for the micromagnetic simulations. a) Magnetization $M$ as a function of the magnetic field $\mu_0 H$, with the field oriented along both the *c*-axis (red trace) and the *ab*-plane (black trace). b) Measured $H/M_c$ ratio versus $M_c^2$ data (black dots). The red line is a linear fit to Eq. 1.

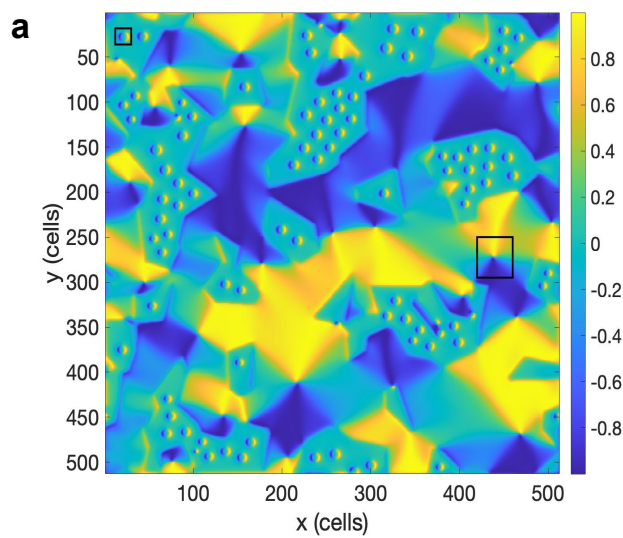

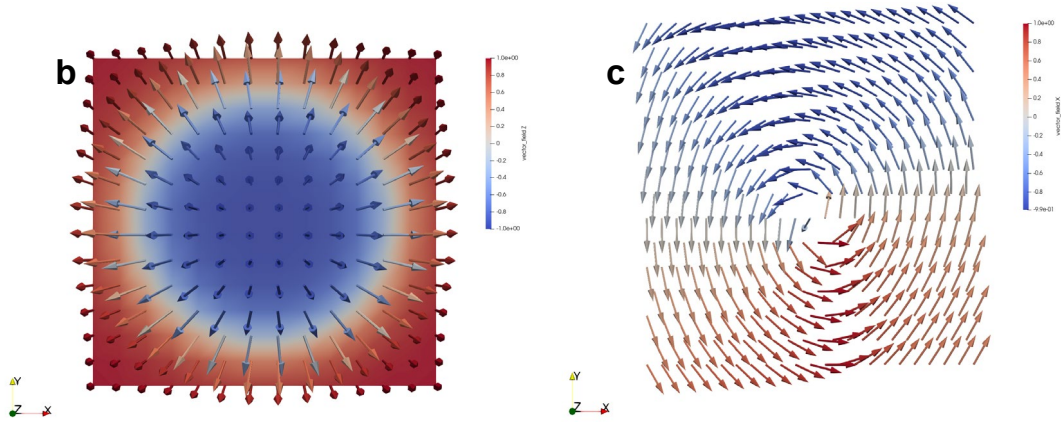

**Figure S11**. Spin textures and Chern numbers. a) Magnetization plot (*x*-component) used to compute the topological charge for skyrmions and merons. b) Isolated skyrmion corresponding to the top left box in a. c) Isolated meron corresponding to the center right box in a. An out-of-plane field of $\mu_0 H = 1600$ G is applied in the $+\hat{z}$ direction with a cell-size of 5 nm.

Calculation of the Topological Charge – 5 nm Resolution

| | |
|---|---|
| 1-Skyrmion corresponding to b | -0.8526 |
| 1-Meron corresponding to c | 0.3894 |
| 16-Skyrmion-Average | -0.8742 |
| 4-Meron-Average | (0.3894244511213203 + 0.4983068948679215 + 0.40951106193104214 + 0.4003776716652857) / 4 ≈ 0.4244 |

**Table S1.** Calculated topological charges using a 5 nm cell. Single and average topological charge for skyrmions and merons, corresponding to Figures S12 a-c. The topological charge was calculated by discretizing **m(r**,t) using the method of finite differences[26].

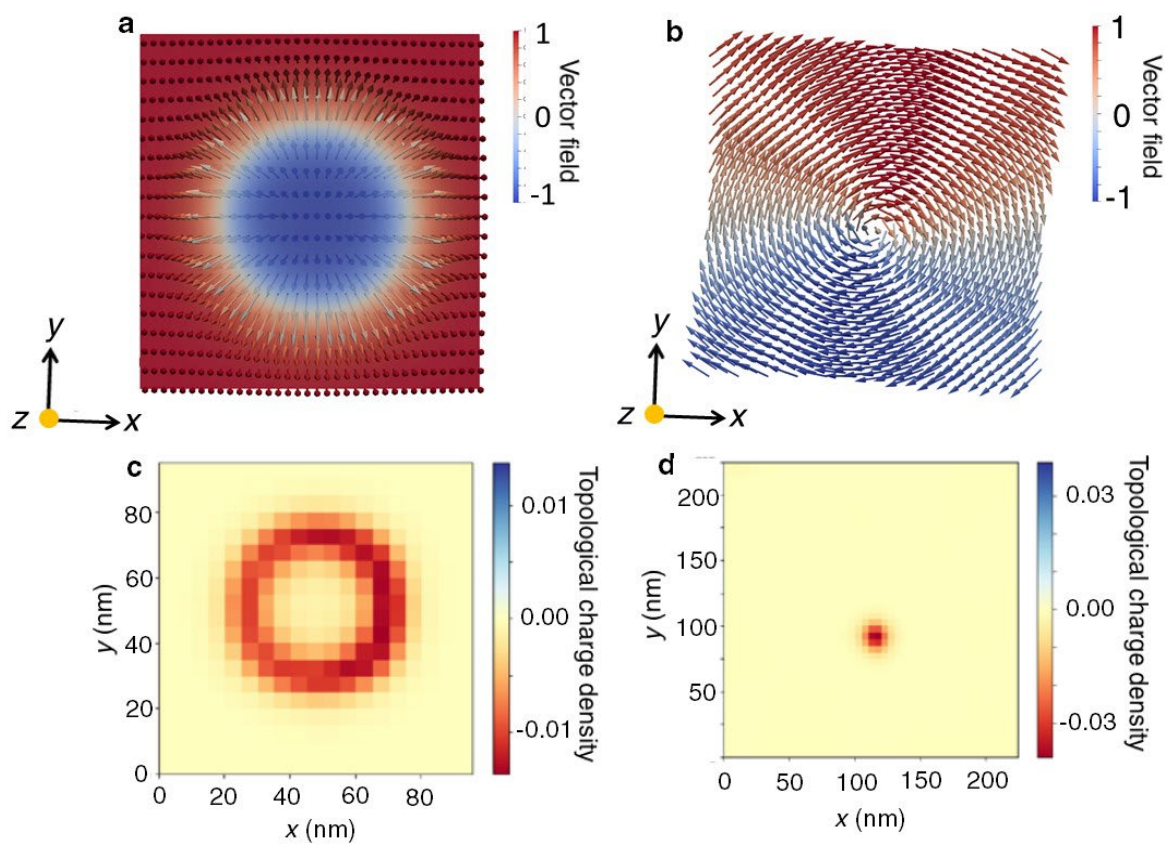

**Figure S12**. Spin textures in a 2 nm cell used for the calculation of the topological charges. a) and b) Same spin textures as in Figure S10 but calculated over a smaller unit cell. c) and d) Respective topological charge densities which, upon integration, yield the topological charges listed in Table S2 below.

Topological Charge - 2nm Resolution

| 1-Skyrmion | -0.9762 |
|---|---|
| 1-Meron | 0.4213 |

**Table S2.** Calculated topological charges using a 2 nm cell. Topological charge for single skyrmion and meron with higher mesh resolution (2nm cell-size), corresponding to Figure S12 a-b. The simulation parameters and relaxation conditions are the same as in Figure S11a.

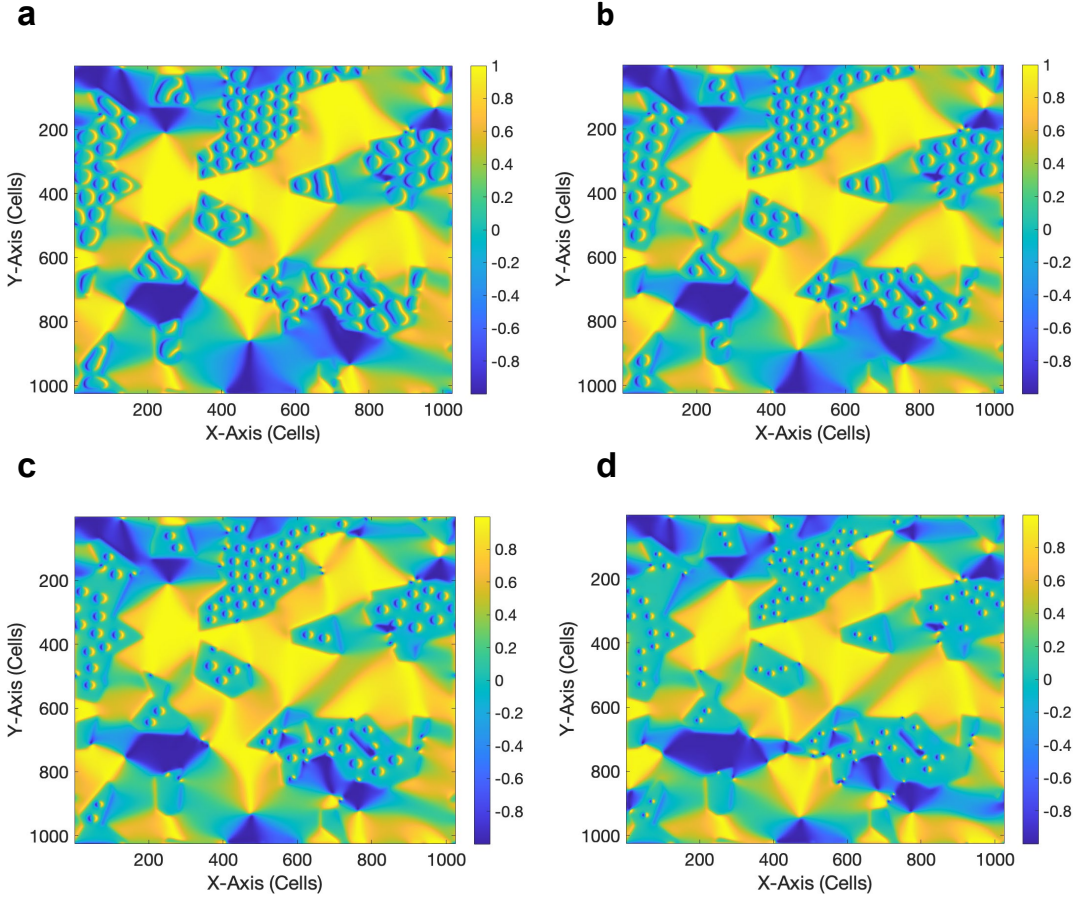

**Figure S13.** Simulated relaxed magnetization state after field cooling and for varying out-of-plane magnetic field values. a) $\mu_0 H = 0$ G, b) 600 G, c) 1300 G and d) 2000 G, showing the shrinkage of skyrmion sizes with increasing magnetic field strength. Simulation parameters are the same as in Figure S11 with cell size of 2 nm.

Methodology

The numerical simulations were performed using the Mumax3 solver[23]. The effective field $\vec{H}_{\text{eff}}$ includes the contribution from perpendicular magnetic anisotropy, Heisenberg exchange, DMI, and an applied external field along the $+\hat{z}$ direction. The DMI was assumed to be purely of interfacial origin, giving rise to the contribution to the energy density [3]

$$\varepsilon = D\left(m_z \frac{\partial m_x}{\partial x} - m_x \frac{\partial m_z}{\partial x} + m_z \frac{\partial m_y}{\partial y} - m_y \frac{\partial m_z}{\partial y}\right)$$

where $D$ is the DMI constant. The first-order magnetocrystalline energy density is given by:

$$\varepsilon_{uniaxial} = -K_{u1}(\boldsymbol{u} \cdot \boldsymbol{m})^2$$

where $K_{u1}$ is the first order uniaxial magnetocrystalline anisotropy constant. The energy density due to the Heisenberg exchange interaction is evaluated as the six nearest neighbour small-angle approximation with energy density given by:

$$\varepsilon_{exch} = -A_{ex}(\nabla \boldsymbol{m})^2$$

where the magnetisation $m$ is taken as the central cell in the nearest-neighbour scheme. The long-range magnetostatic field is evaluated as a discrete convolution of the magnetization with a demagnetizing field kernel $K$

$$B_{demag\ i} = K_{ij} * m_j$$

where $M = M_s m$ is the unnormalized magnetization, with $M_s$ the saturation magnetization (A/m). The corresponding energy density is provided by:

$$\varepsilon_{demag} = -\frac{1}{2} M \cdot B_{demag}$$

Discussion

Regions of in-plane and out-of-plane anisotropy comprised of Thiessen polygons were generated with Voronoi tessellation, using a grain size of 200 nm. The material parameters are exchange constant $A = 10$ pJ m$^{-1}$, Gilbert damping $\alpha = 0.3$, saturation magnetisation $M_s = 630$ kA m$^{-1}$ along $c$ plane, $M_s = 730$ kA m$^{-1}$ along $ab$ plane, $D = 1.2$ mJ m$^{-2}$ and $K_u = 2.5$ kJ m$^{-3}$ in the out-of-plane regions, and periodic boundary conditions were applied in the lateral film dimensions. The magnetization was initially randomized before relaxing the magnetic material in the presence of a $+\hat{z}$ directed magnetic field. In Figure S13, the field was reduced from 2000 G to 0 G in steps of ~633 G along the $+\hat{z}$ direction.

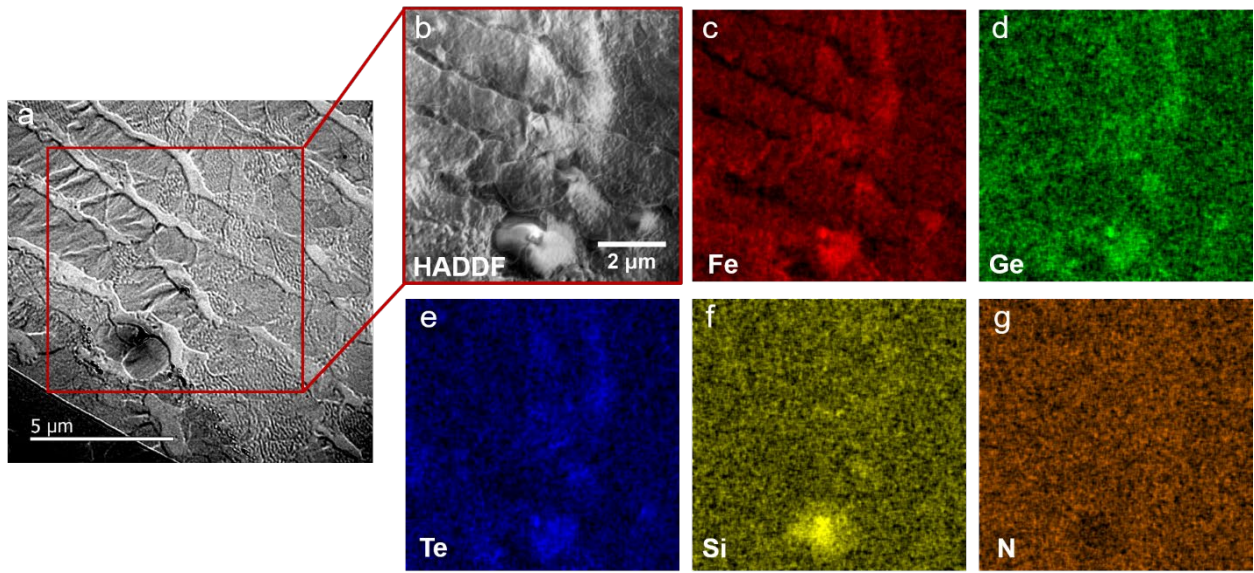

**Figure S14.** Energy dispersive spectroscopy analysis of a Fe$_{5-x}$GeTe$_2$. a) LTEM image of the exfoliated Fe$_{5-x}$GeTe$_2$ single-crystal. b) High-angle annular dark-field scanning transmission electron microscopy (HAADF-STEM) image of a particular region of an exfoliated Fe$_{5-x}$GeTe$_2$ single-crystal indicated by the red box in a. c), d), and e) Energy dispersive spectroscopy (EDS) mappings indicating the distributions in Fe, Ge, and Te respectively. EDS yields average concentrations of 5, 1, and 2.3 for Fe, Ge, and Te, respectively. e) Transmission electron microscopy image of the same area, where one also incorporates the contributions: f) silicon, g) nitrogen to the EDS analysis. The incorporation of these elements to the EDS analysis yields the average ratios of 4.99, 1, and 2.2 for Fe, Ge, and Te, respectively. In summary, Conventional EDS analysis yields variations in the respective atomic contents in the order of 1 to 5 %, which is within, or close to, the typical error bars of the EDS technique. The STEM-EDS measurement was performed using FEI Talos F200X TEM at 200 kV equipped with a super X energy-dispersive spectrometer from Bruker.

6. Calculation of emergent fields and associated Hall response.

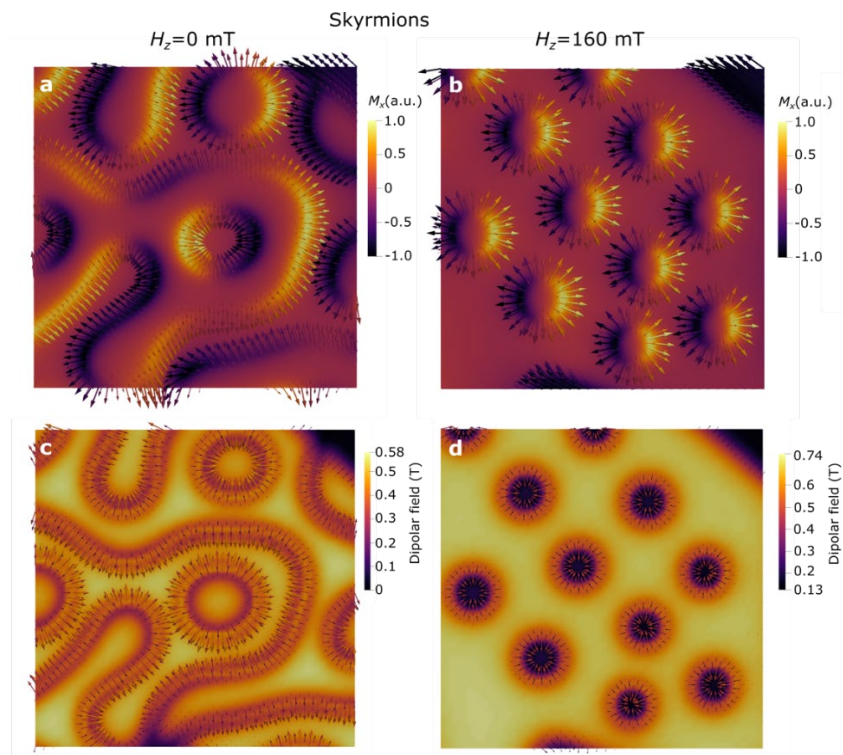

**Figure S15.** a,b) Snapshots of the in-plane component of the magnetization ($M_x$) obtained via micromagnetic simulations for a perpendicular field $H_z$ of 0 mT and 160 mT, respectively. c,d) Dipolar fields calculated for configurations a-b, respectively. Only areas contained skyrmions have been included here.

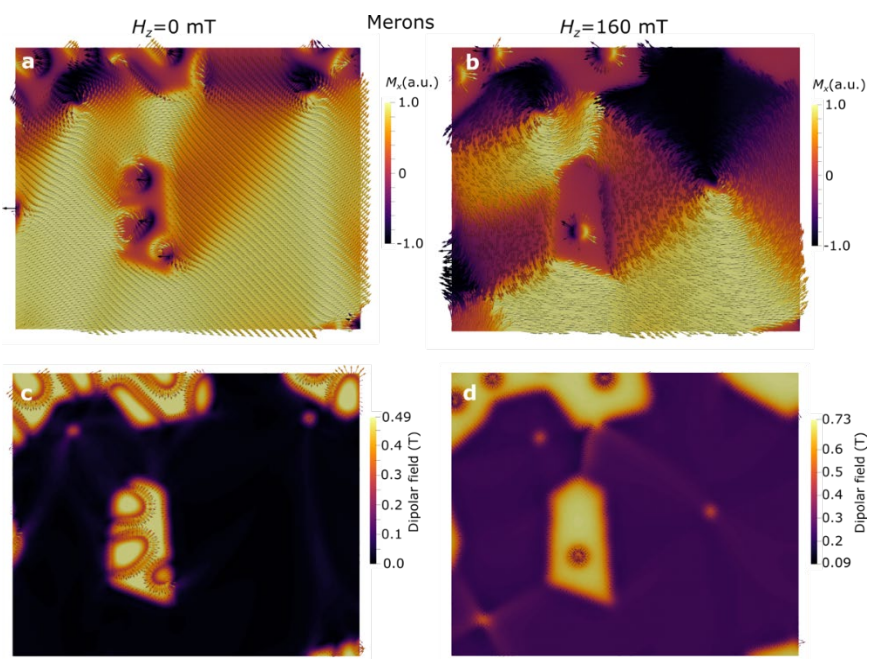

**Figure S16.** a,b) Snapshots of the in-plane component of the magnetization ($M_x$) obtained via micromagnetic simulations for a perpendicular field $H_z$ of 0 mT and 160 mT, respectively. c,d) Dipolar fields calculated for configurations a-b, respectively. Only areas containing merons have been included here.

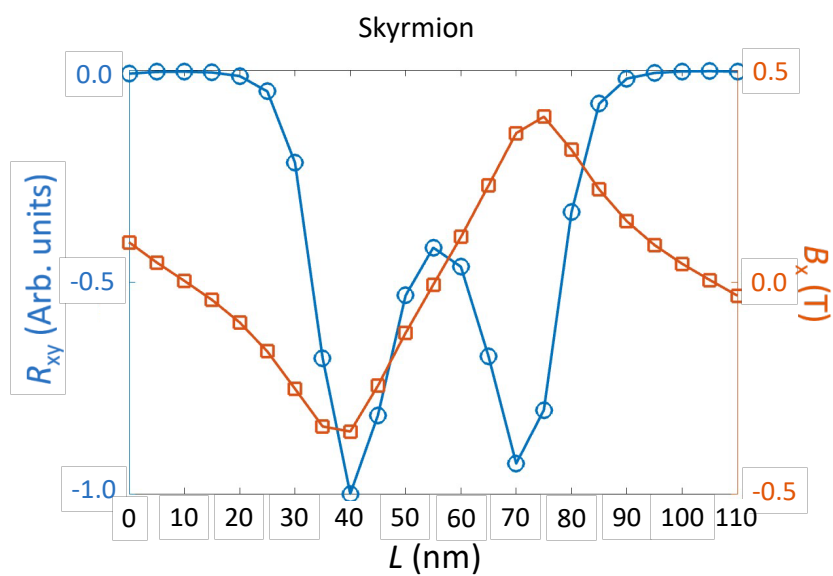

**Figure S17**. Calculated profiles for the Hall resistance $R_{xy}$ (right) and the emergent magnetic field $B_x$ (left) across a skyrmion in $Fe_{5-x}GeTe_2$. The distance $L$ measures the length considered for the computation.

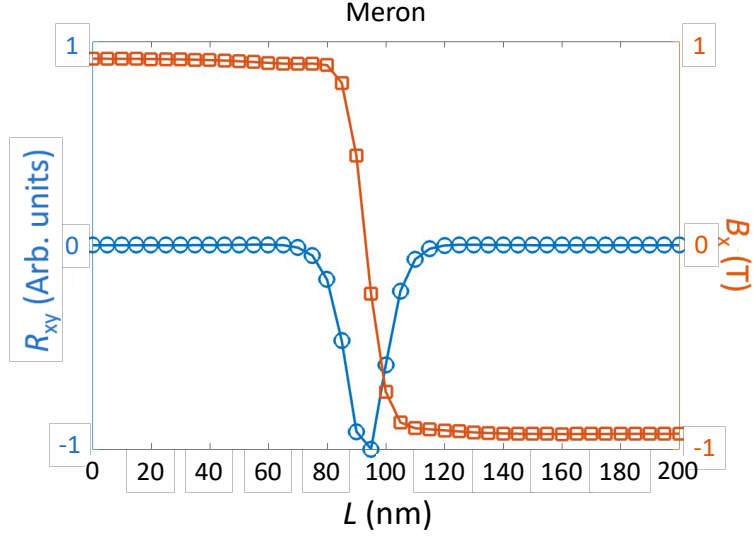

**Figure S18**. Calculated profiles for the Hall resistance $R_{xy}$ (right) and the emergent magnetic field $B_x$ (left) across merons in Fe$_{5-x}$GeTe$_2$. The distance $L$ measures the length considered for the computation.

6. Determination of the domain wall width between planar domains.

A magnetic domain wall separating in-plane domains having a relative orientation of 180° and characterized by domain wall width $\delta_{dw}$, can be described by a hyperbolic tangent function[35]:

$$B_y = a + b \tanh\left(\frac{x-c}{\delta_{dw}}\right) \quad (2)$$

where $a$, $b$, and $c$ are constants. We can estimate $\delta_{dw}$ by fitting the profile of the magnetic induction along $y$-axis to Eq. 2. Figure S15a shows an experimental map of the $y$-component of the magnetic induction map of Fe$_{5-x}$GeTe$_2$. For frame 1 the fitting results in a line profile that is presented in Figure S15b yielding a value for $\delta_{dw}$ of about $(26.5 \pm 0.3)$ nm. However, as the original Lorentz image leading to Figure 9 a) was collected through an out-of-focus condition, this estimated value would be incorrect. Therefore, a through-focus-series of Lorentz images were recorded for a precise estimation of $\delta_{dw}$. Figure S15c displays the experimental domain wall width as a function of the defocus length df, leading to a more precise value for $\delta_{dw}$ at df = 0 mm. The evaluated value of $\delta_{dw}$ is found to be 20.9 nm for region 1. We computed the mean values of $\delta_{dw}$ over the four regions, enclosed by the red frames in Figure S15a, obtaining an average $\delta_{dw}$ value of $(25 \pm 5)$ nm.

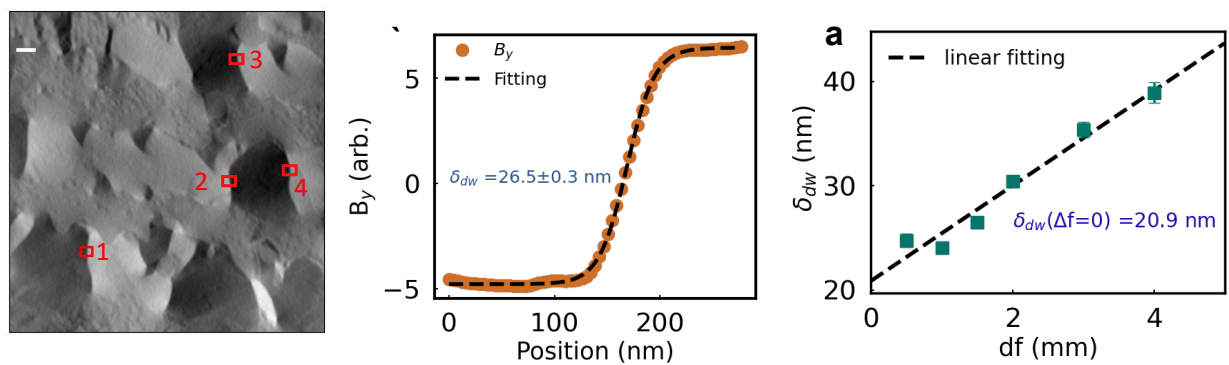

**Figure S19**. Determination of the domain wall width through Lorentz TEM. a) *y*-component integral of the magnetic induction map of a Lorentz TEM image collected with the defocus length of 1.5 mm. Four red frames indicate highlighted regions used for the estimation of domain wall width. b) Line profile, averaged along the height direction, across two neighboring domains with a 180° relative orientation as marked by frame 1 in a. c) Fitted domain wall width $\delta_{dw}$ as function of defocus length df. Dashed line is a linear fit to Eq. 2.